\documentclass{JHEP3}

\usepackage{graphicx}

\def\be{\begin{equation}}
\def\ee{\end{equation}}
\def\bea{\begin{eqnarray}}
\def\eea{\end{eqnarray}}
\def\ba{\begin{array}}
\def\ea{\end{array}}
\def\eq{\begin{equation}}
\def\eqe{\end{equation}}
\def\eqa{\begin{eqnarray}}
\def\eqae{\end{eqnarray}}
\def\ena{\end{eqnarray}}

\def\ie{{\it i.e.~}}
\def\nn{\nonumber}
\def\Tr{{\rm Tr}}

\def\bbR{{\mathbb R}}
\def\bbZ{{\mathbb Z}}
\def\ra{\rangle}
\def\la{\langle}
\def\N4{{\cal N}=4}
\def\AdSS5{$AdS_5$}
\def\AdS5s5{$AdS_5 \times S^5$}
\def\AdSs5{$AdS_5$}
\def\AdS5s5{$AdS_5 \times S^5$}

\def\gy{g_{_{YM}}}

\def\calP{{\cal P}}

\def\calN{{\cal N}}
\def\calC{{\cal C}}

\def\calP{{\cal P}}

\def\calF{{\cal F}}

\def\calA{{\cal A}}

\def\calQ{{\cal Q}}

\def\calL{{\cal L}}
\def\calD{{\cal D}}

\def\a{{\alpha}}

\def\adot{{\dot\alpha}}

\def\c1{{\chi^1}}

\def\t2{\tau_2}

\newbox\SlashedBox
\def\fs#1{\setbox\SlashedBox=\hbox{#1}
\hbox to 0pt{\hbox to 1\wd\SlashedBox{\hfil/\hfil}\hss}{#1}}
\def\hboxtosizeof#1#2{\setbox\SlashedBox=\hbox{#1}
\hbox to 1\wd\SlashedBox{#2}}

\def\ms#1{\setbox\SlashedBox=\hbox{$#1$}
\hbox to 0pt{\hbox to 1\wd\SlashedBox{\hfil/\hfil}\hss}#1}

\def\IZ{\relax\ifmmode\mathchoice
{\hbox{\cmss Z\kern-.4em Z}}{\hbox{\cmss Z\kern-.4em Z}}
{\lower.9pt\hbox{\cmsss Z\kern-.4em Z}} {\lower1.2pt\hbox{\cmsss
Z\kern-.4em Z}}\else{\cmss Z\kern-.4em Z}\fi}

\title {Instanton corrections to  circular Wilson loops\hfill\break  in
${\cal N}$=4 Supersymmetric Yang--Mills}

\author{Massimo Bianchi \\ Dipartimento di Fisica, \ Universit{\`a} di Roma \
``Tor Vergata'', Via della Ricerca  Scientifica, 1,
00133 \ Roma, \ ITALY \\
E-mail: {\tt Massimo.Bianchi@roma2.infn.it},}
\author{Michael B. Green and Stefano Kovacs \\
Department of Applied Mathematics and Theoretical Physics \\
Wilberforce Road, Cambridge CB3 0WA, UK \\
E-mail:
\email{M.B.Green@damtp.cam.ac.uk},
\email{S.Kovacs@damtp.cam.ac.uk}}

\abstract{It is argued that whereas supersymmetry requires
the instanton contribution to the expectation value of a straight
Wilson line in  the ${\cal N}=4$  supersymmetric $SU(2)$ Yang--Mills
theory to vanish, it is not required to vanish
in the case of a circular Wilson loop.   A non-vanishing value can arise
from a subtle interplay between a divergent integral over bosonic
moduli and a vanishing integral over fermionic moduli. The
one-instanton contribution to such Wilson loops
is explicitly evaluated in semi-classical approximation.  
The method utilizes the
symmetries of the problem to perform the integration
over the bosonic and fermionic collective coordinates of the
instanton.  The integral is singular for small instantons touching the loop
and is regularized by introducing a cutoff  at the boundary of the
(euclidean) $AdS_5$ moduli space.  In the case of a circular loop
a nonzero finite result is
obtained when the cutoff is removed  and  a perimeter divergence
subtracted. This is contrasted with  the case of the straight line
where the result is  zero after subtraction of an identical
divergence per unit length.  The linear divergence is an
artifact of our non supersymmetric regulator that deserves further
consideration.   The generalization to  gauge group
$SU(N)$ with arbitrary $N$ is straightforward in the limit of small
't Hooft coupling.  The extension to strong 't Hooft coupling is more
challenging and only  a qualitative  discussion is given of the
AdS/CFT correspondence.}

\preprint{DAMTP-2001-59; ROM2F/2001/28; hep-th/0202003}

\begin{document}
\section{Introduction}
\label{intro}

The correspondence between ${\cal N}=4$ supersymmetric $SU(N)$
Yang-Mills theory and type IIB superstring theory on
$AdS_5\times S^5$ has been the subject of extensive study and is by
now well tested in the large $N$ limit (see for instance
\cite{magoo,dan,mb,dfedh}). Many of these tests involve  comparison of
correlation functions of gauge-invariant Yang--Mills operators with
corresponding amplitudes in type IIB supergravity (the small
$\alpha'/\ell^2$ limit of superstring theory in $AdS_5\times S^5$,
where $\ell^{2}$ is the scale of the curvature).  By comparison, the
correspondence involving nonlocal Yang--Mills operators, such as
Wilson loops, has been  relatively little studied. This is one of the
main motivations for this paper.

The Wilson loop in pure Yang--Mills theory
is the expectation value of the holonomy,
$\la W({\cal C}) \ra = \la \Tr\, \calP \exp i\, \int_\calC
A_\mu dx^\mu \ra $, which is a functional of an arbitrary curve ${\cal
C}$.  It is   of central importance
since it is an order parameter that characterizes the different
phases of the theory.  When the trace is taken in the fundamental
representation of the gauge group $G$, such as the $N$-dimensional
representation of $SU(N)$, the Wilson loop decreases as the
exponential of the area in a confining phase but as the exponential of
the perimeter in a non-confining phase.  A long rectangular loop
 determines the static potential
between elementary test charges, such as quark charges in QCD.
Similar considerations apply to Wilson loops in more general gauge
theories in which there are additional dynamical fields in the adjoint
representation.  In supersymmetric theories the concept of the gauge
connection generalizes to a superfield that contains other components
beyond the usual vector potential.  Correspondingly, the Wilson loop
generalizes in a natural manner to include contributions from the
extra fields.  The ${\cal N}=4$ supersymmetric Yang--Mills theory is
worthy of study in its own right since it is an example of a
nontrivial superconformal field theory in four dimensions
and a prototype for more
general and realistic gauge theories.  The natural generalization of
the Wilson loop in this theory is defined by
\be
\la W({\cal C}) \ra = {1\over N} \langle {\rm Tr} \, {\cal P} \exp
{\{i\int_{{\cal C}} (A_{\mu} \dot{x}^{\mu} + i \varphi_{i} \dot
y^i + [\bar\theta_{A}\dot{x}_{\mu} \sigma^{\mu} \lambda^{A} +
\theta^{A}\dot{y}_{i} {\widehat\Gamma}^{i}_{AB} \lambda^{B} + {\rm
h.c.}] +  \cdots ) ds\} } \rangle \: . \label{wildef}
\ee
where $\lambda^A$ ($A=1,2,3,4$ indicates a ${\bf 4}$ of the R-symmetry
group, $SU(4)$) and $\varphi_i$ ($i=1,\dots,6$ labels the ${\bf 6}$
of $SU(4)$)  are the fermion and scalar fields in the ${\cal N}=4$
supermultiplet.   The  matrices ${\widehat\Gamma}^i_{AB}$ and
$\bar{\widehat\Gamma} ^{i\,  {AB}}$ are $SO(6)$ Clebsch--Gordan
coefficients that couple a {\bf 6}  to two ${\bf 4}$'s and to two
${\bf \bar 4}$'s, respectively.  The curve $\calC$ now represents the curve
in `superspace' --- in  other words this kind of  Wilson  loop depends
not only on the curve $x^\mu(s)$ but also on six
extra variables $y^i(s)$  and on
the spinors $\theta^A(s)$  and  $\bar \theta_A(s)$ that contain the
sixteen odd  (Grassmann) variables  of ${\cal N}=4$ on-shell
superspace. The $\cdots$ in (\ref{wildef}) stands for terms involving
higher powers of the fermionic coordinates, $\theta^A$ and
$\bar\theta_A$.  The expression (\ref{wildef})  is appropriate to
euclidean signature whereas the factor of $i$ in the coefficient of
 $\varphi_i$ is absent with Minkowski signature.  Its presence  is
important, among other reasons, because  it implies that the
exponential is not purely a phase.  This  expression  can be
motivated in various ways.  For example, the loop can be considered to
be the holonomy of  an infinitely massive $W$-boson that is generated
by  breaking the  gauge group $SU(N+1)$ to $SU(N)\times U(1)$  (as
shown in an appendix of \cite{dgo}).

Wilson loops of this kind were first studied in \cite{mald,rey}, where
their interpretation within the AdS/CFT  correspondence was stressed.
Within string theory  the Wilson loop is  interpreted as the
functional integral over all world-sheets embedded  in $AdS_5$ and
bounded by the loop.  In the supergravity limit (the  small $\alpha'$
limit of string theory) this integration over  fluctuating
surfaces is dominated by the  surface of minimum area  $A_{\rm min}$ in
$AdS_5$.  The behaviour of the loop is therefore $\la W(\calC) \ra \sim
\exp(- A_{\rm min})$.  Since the metric is singular near the  boundary of
$AdS_5$ an infinite perimeter term arises, representing the mass of
the  test particle circulating in the loop (it was shown in \cite{dgo}
that this divergence may be eliminated by an appropriate choice of
world-sheet boundary conditions).  The finite
minimal area obtained by subtracting the divergent piece is a unique
and well defined quantity \cite{graham}.

As shown in \cite{orrt} the expression (\ref{wildef}) is invariant
under $\kappa$ transformations of the one-dimensional theory on the
test particle world-line,
provided the curve satisfies appropriate
conditions and the Yang--Mills fields satisfy their equations of
motion.  This is closely related to the $\kappa$ symmetry
of the massless ($p^2=0$) ten-dimensional superparticle.  A standard
argument based on
gauge-fixing of $\kappa$ symmetry
then implies that the loop is invariant under half of the 32
superconformal supersymmetries.  In that case the supersymmetries
are defined by  spinor parameters  $\kappa^A_\alpha$,
$\bar\kappa_A^{\dot \alpha}$ that are related by
\be
\dot{x}_{\mu}\, \sigma^{\mu}\, \bar\kappa_{A} = \dot y_i\,
{{\widehat\Gamma}}^{i}_{AB} \kappa^{B} \: .
\label{relloop}
\ee
It is easy to see that this constrains $\dot y^i$ so  that
\be
\dot y^i = n^i |\dot x|,
\label{condit}
\ee
where $n^i$ is an arbitrary fixed unit vector on the five-sphere
($n^2=1$).   As we will see later
the solutions to (\ref{relloop}) have the form\footnote{The subscript 
${}_{_{\oplus}}$ is used to
label the parameters to avoid later confusion with the instanton
moduli.} $\kappa^A_\alpha =
\eta_{{_\oplus}\, \alpha}^A +(\sigma\cdot x
\bar\xi_{{_\oplus}}^A)_{\alpha}$, $\bar \kappa_A^{\dot \alpha} =
\bar\eta_{{_\oplus}\,A}^\adot +(\bar\sigma\cdot x
\xi_{{_\oplus}\,A})^{\adot}$, where the sixteen Poincar\'e supersymmetry
parameters,
$\eta^A_{_{\oplus}\, \alpha}$ and $\bar\eta_{_{\oplus}\, A}^\adot$,
and the sixteen conformal supersymmetry parameters,
$\bar\xi^A_{_{\oplus}\, \adot}$ and $\xi_{_{\oplus}\, A}^\alpha$,  are
related.   We will only  consider the special loops
in which we set $\theta^{A}(s) = 0$ (so the terms in (\ref{wildef})
that depend on $\theta$ are absent).
A particularly symmetric example of a Wilson loop
satisfying (\ref{relloop}) is the circular loop of radius $R$.
Superconformal invariance implies that the expectation value of such a
loop cannot depend on $R$ so that $\la W \ra_{\rm circle}$ is a constant,
but it does depend in a non-trivial way on the dimensionless
parameters $\gy$ and $N$.  The special feature of  a circular loop is
that
(\ref{relloop}) implies that an {\it $x$-independent} linear
combination of
Poincar\'e and conformal supersymmetries remains unbroken.
In other words, the sixteen unbroken supersymmetries are global whereas they
are $x$-dependent for a generic loop satisfying  (\ref{relloop}).
However, in the quantum theory it is necessary to introduce a cut
off.  As we will see later, this necessarily breaks the remaining
supersymmetries and such Wilson loops receive quantum corrections.
A class of perturbative contributions to the expectation values of
Wilson loops of this kind has been calculated to all orders in the
coupling constant and it has been argued that it contains all the
relevant contributions, at least in
the large $N$ limit \cite{esz}.  This consists of the `rainbow
diagrams' --- the class of planar diagrams in which all propagators
begin and end on the loop (so there are no internal interaction
vertices). The sum of such diagrams was determined in terms of a
zero-dimensional gaussian matrix model. A suggestion has been made
\cite{dg} for  extending this to all orders in the $1/N$ expansion (as
well as all orders in the 't Hooft coupling) by
use of  an anomaly argument.  This was arrived at  by considering a `straight'
Wilson line on $\bbR^4$.  This case is even more special since
(\ref{relloop}) now implies that a subset of
the Poincar\'e supersymmetries are unbroken and do not mix with the
conformal supersymmetries.  The unbroken Poincar\'e supersymmetries
(but not the conformal supersymmetries) are
preserved in the quantum
theory in the presence of a suitable cutoff
and protect the Wilson line expectation value so that
$\langle W\rangle_{\rm line} =1$. A circular loop passing through the
origin is mapped to a straight line by a conformal inversion and it is
the
associated conformal anomaly that  gives rise to  a nontrivial expression
for $\la W \ra_{\rm circle}$ as a function of the coupling. The argument in
\cite{dg} suggested that the gaussian matrix model results should be
taken seriously for all values of $N$, not simply in the large-$N$
limit contemplated by \cite{esz}.  However, as shown in \cite{ad}
there is a wide class of matrix models with nontrivial potentials
which give rise to the same leading $N$ behaviour but give different
expressions for the Wilson loop at finite $N$.  Recent computations
\cite{plefka} have further questioned  the validity of the conjecture
within
perturbation theory.  From our perspective, it is notable that the
expression for the
circular Wilson loop suggested by \cite{dg} has no instanton
contributions and does not depend on the vacuum angle, $\vartheta$.

 In this paper we will explicitly compute the one-instanton
contribution to a
circular Wilson loop  in $SU(2)$ $\N4$ Yang--Mills  in semi-classical
approximation -- to lowest order in the Yang--Mills coupling constant,
$g_{_{YM}}$. A preliminary outline of this work was presented in
\cite{bgk}.

\subsection{Expectations based on supersymmetry}
\label{expect}

Before carrying out the calculations in detail it is of interest to use
the symmetries of the problem to anticipate the result\footnote{We are
grateful to Juan Maldacena for conversations on the following points.}.  It is
instructive to contrast the expression for the one-instanton contribution
to the circular  Wilson loop with that of the  `straight line' (of the
kind considered in \cite{dg}).  Naively (ignoring the need for a
cutoff), the loop preserves half of the  Poincar\'e and conformal
supersymmetries.  At least some of these are broken by the presence of
an instanton.  For every
supersymmetry of the background
that is broken by the instanton there is a `true' fermionic
modulus (a fermionic integration variable that does not enter
into the integrand) so the integration over supermoduli space
vanishes.  However,
this argument is too naive since the integration over the bosonic
moduli, which parameterize $AdS_5$, diverges on the boundary, \ie for
instantons of small scale size.  It is therefore essential to
introduce a cutoff.  In principle, such a cutoff can be introduced by
considering the $SU(N)$ theory as the limit of a $SU(N+1)$ theory
spontaneously broken to $SU(N) \times U(1)$, in which the scalar field
vacuum expectation value, $M$, becomes infinite \cite{dgo}. The
$W$-bosons have mass $M$ and are in the $N$, $\bar N$ of $SU(N)$.  In
the limit $M\to \infty$ the Wilson loop can be defined in terms of the
holonomy of a $W$-boson with a specified trajectory.  The Wilson loop
can be regulated by keeping the mass, $M$, finite but large (compared
to the inverse radius of the loop)  in much the same way as considered
in \cite{mald}.  In this case the fluctuations of the test particle
are non-zero and the loop is smeared out over a region $M^{-1}$.

In the absence of the loop the cut-off theory preserves the sixteen
Poincar\'e supersymmetries (with parameters $\eta_{_{\oplus}},
\bar\eta_{_{\oplus}})$ but breaks the sixteen conformal
supersymmetries (with parameters $\xi_{_{\oplus}},
\bar\xi_{_{\oplus}}$), since the cut-off theory is not conformally
invariant.  We now want to consider the cut-off theory in the presence
of the Wilson loop, which breaks further supersymmetries, as
determined by the condition (\ref{relloop}).  It is easy to see from
this equation that in the case of a straight line ($\dot x_\mu \propto
x_\mu$) the Poincar\'e supersymmetries of opposite chiralities are
related to each other ($\eta_{_{\oplus}}$ is related to
$\bar\eta_{_{\oplus}}$).  Therefore, there are eight residual
supersymmetries of the Wilson line background in the cut-off theory.
In the case of
the circular loop the condition (\ref{relloop}) relates the Poincar\'e
supersymmetries to the conformal supersymmetries.  Since the conformal
supersymmetries are already broken by the cutoff we see that a
circular loop in the cut-off theory does not preserve any
supersymmetry.  Later, we will also consider a cutoff that preserves
a $SO(5)$ symmetry instead of  Poincar\'e symmetry.  Such a cutoff is
very natural when considering circular loops on $S^4$, which is
convenient for displaying conformal symmetry.  We will argue that a
$SO(5)$-invariant cutoff cannot preserve any supersymmetry,  even in
the absence of the loop.

 Now consider the introduction of the instanton.  In the case of the
straight line with the
Poincar\'e-invariant cutoff the eight unbroken supersymmetries are
broken by the instanton.  This generates eight true fermionic moduli
so the integrated expression for the instanton contribution should
vanish.  More generally, the presence of unbroken supersymmetries with
this cutoff requires $\la W \ra_{\rm line} =1$, as seen in the perturbative
sector \cite{esz}, \cite{dg} and in line with expectations based on
the AdS/CFT correspondence \cite {mald}.  In the case of the circular
Wilson loop there are no surviving supersymmetries to be broken in the
presence of the  cutoff  (whether it is Poincar\'e invariant or
$SO(5)$ invariant) so we conclude that the instanton contribution to
the  loop expectation value can be nonzero and must be independent of the
cutoff.

In the following we will make use of the standard BPST instanton
solution of the Yang--Mills equations.  In principle, these equations
should be modified to include the effect of the Wilson loop source,
which changes the standard BPST instanton solution of the sourceless
theory.  However, at least when the instanton is not too singular,
the corrections to the equations induced by the
current source are suppressed by powers of $g_{_{YM}}$ so they can be
neglected in the  semi-classical approximation. The singular configuration,
in which a small instanton touches the loop, should be regulated in the
Poincar\'e-invariant manner described above.  In practice, we will regulate
the singularity by introducing a cutoff in the integral over the
collective coordinates of the instanton near the boundary of $AdS_5$.
This cutoff breaks supersymmetry and potentially introduces an ambiguity
in the finite value of the expectation value of the circular Wilson 
loop which will be discussed in the last section.

\subsection{Strategy and layout}
\label{layout}

As usual, the  instanton computation boils down to an integral over
the supermoduli space spanned by eight bosonic and sixteen fermionic
collective coordinates. The bosonic collective coordinates correspond 
to broken translation, scale and gauge symmetries and will be denoted 
by $( x_{0}^{\mu}, \rho_{0}, \alpha^{a}_{0})$, respectively 
(although the gauge moduli  will be irrelevant in the
following).  The fermionic collective coordinates are associated with 
broken Poincar\'e supersymmetries and conformal supersymmetries and 
will be denoted by $(\eta^{A}, \bar\xi^{A})$, respectively.
The prospect of directly evaluating even the bosonic
part of the Wilson loop is somewhat daunting.  However, considerable
simplification arises after taking advantage of the conformal
symmetries of this system.  The presence of the loop breaks the
$SO(4,2)$ conformal invariance but, as we will show in section
\ref{symmetries}, for a circular loop there is a residual unbroken
$SO(2,2)=SO(2,1)\times SO(2,1)$ subgroup.  When considering the
instanton calculation we will be interested in the euclidean version
of the theory in which the $AdS_5$  boundary is $S^4$.  The euclidean
conformal group is $SO(5,1)$ and the subgroup that leaves the loop
invariant is $SO(3)\times SO(2,1)$.  We will further show that the  presence of a
circular loop in the ${\cal N}=4$ supersymmetric theory breaks the
full $SU(2,2|4)$ superconformal symmetry to a residual $OSp(2,2|4)$,
which has sixteen fermionic generators.

These symmetries will be used to determine the structure of the
instanton contribution to the Wilson loop in a toy bosonic model in
section \ref{pureYM}.  This calculation  includes only the bosonic
moduli  of the complete $\calN=4$ Wilson loop calculation (and is {\it
not} the same as the expression for the instanton contribution to a
Wilson loop in pure Yang--Mills theory, which has a non 
conformally-invariant 
measure\footnote{The complete calculation in the Yang--Mills
theory, even to lowest nontrivial order in the coupling, is
significantly more subtle.}).  In order to streamline the discussion of the
conformal properties it will prove convenient to make use of
Dirac's formalism \cite{dirac} for representing the conformal group by
extending four-dimensional Minkowski space-time  to six dimensions
with signature $(4,2)$ with coordinates $X_M$ ($M=0,\dots,5$), where
$X_0$ and $X_4$ are time-like (and $X_0$ will be Wick rotated when
describing the euclidean theory). In this way, the $SO(4,2)$ and
$SO(2,2)$ symmetries can be represented linearly by rotations (and
boosts) on $X_M$.  Imposing the invariant constraint  $X^MX_M=\ell^2$,
where $\ell$ is an arbitrary dimensional constant, results in a
representation of $SO(4,2)$ in $AdS_5$ and thence to a
four-dimensional representation on the boundary of $AdS_5$.  This will
be reviewed in section \ref{dirac6dim} and appendix \ref{conformal}.

The coordinates of euclidean $AdS_5$ (which is the ball, $B^5$) enter
as collective coordinates in the instanton problem.  The fact that the instanton is
invariant in form under arbitrary  conformal transformations, up to an irrelevant
gauge transformation, together with the invariance of the loop under
$SO(3) \times SO(2,1)$, will be used to evaluate the Wilson loop integrand.
The expression for the integrand of the Wilson loop with an instanton at a
generic point in moduli space, $(x^\mu_0, \rho_0)$, 
is identical to that in which the instanton is located at any
point on the same $SO(3) \times SO(2,1)$ orbit.  In particular, it is the same
as if the instanton were at the
centre of the loop, $\tilde x_0^\mu=0$ with a scale $\tilde
\rho(x_0,\rho_0)$ that is a certain $SO(3)\times SO(2,1)$-invariant
function of $X_M$.   But the expression for the
Wilson loop with an instanton at the centre reduces to one
with an abelian gauge field, in which case the
path ordering is trivial which makes the integration over the
moduli very simple.   The integral is very
divergent since there is no suppression of instantons located arbitrarily far from the
loop.  This type of divergence  does not appear in the $\calN =4$ supersymmetric case.
However,  there is also a divergence from small scale instantons touching the loop, which
also needs to be addressed in the supersymmetric case.
All of these divergences are regulated by  imposing  an
$SO(5)$-invariant cutoff on the integration over the bosonic moduli
that excludes a small spherical shell close to the boundary of the
moduli space, $AdS_5$.

 In section \ref{lineambiguity} we will
consider the expression for the instanton contribution in the bosonic
toy model to the straight
Wilson line, which is defined on $\bbR^4$.
This was the starting  configuration considered in \cite{dg}.  In the
case of a  straight line it is  natural to use a Poincar\'e invariant   cutoff,
$\rho_0 \ge \epsilon$ (where $\epsilon$ is an infinitesimal constant),
which is invariant under the translational isometry of the straight
line.   Both cutoffs (indeed, all possible cutoffs) break conformal
invariance.   The  distinction between the $SO(5)$-invariant and
Poincar\'e invariant cutoffs is crucial in determining the difference
between the expectation values of the circular Wilson loop and straight Wilson
line  in the superconformal theory
considered later.

In sections \ref{instsupsp} and \ref{N4oneinst} we will describe the
extension of these ideas to the ${\cal N}=4$ supersymmetric case in
which there are sixteen fermionic moduli in addition to the bosonic
moduli. The  bosonic fields $\varphi^i$ and $A_\mu$ have zero modes
that are  induced by the couplings to the fermionic sources in the
usual manner. The
multiplet of these zero modes can be generated from the classical
instanton profile of the vector potential by successive application of
the supersymmetries that are broken by the instanton.  This leads to
an expression for  $\varphi$ that is polynomial in fermionic moduli,
beginning with a quadratic term
\be
\hat \varphi^{ia} =
{1\over 2} F^{a-}_{\mu\nu} {{\widehat\Gamma}}^{i}_{AB} \zeta^{A}
\sigma^{\mu\nu} \zeta^{B} + \cdots \: ,
\label{vardef}
\ee
where the  hat  indicates the value of a field containing fermionic
collective coordinates induced by the instanton background.  In this
expression the fermionic moduli are  packaged into the chiral spinor
\be
\zeta^{A} (x) = \eta^{A} + x_{\mu} \sigma^{\mu} \bar\xi^{A} \: ,
\label{zetdef}
\ee
which, up to rescalings, is also a   $(1-\gamma_5)$ projection of a
Killing spinor of $AdS_5$ and indicates  the holographic connection
between the Yang--Mills instanton and the D-instanton of the IIB
string theory in $AdS_5\times S^5$ \cite{bg,bgkr}. The anti self-dual
field strength of the instanton in (\ref{vardef}) is given by
\be
F^{a-}_{\mu\nu} = {4
\eta^{a}_{\mu\nu} \rho_{0}^{2} \over ((x-x_{0})^{2} +
\rho_{0}^{2})^{2} } \: ,
\label{fielddef}
\ee
where  $\eta_{\mu\nu}^a$ is the conventional 't Hooft symbol.  The
$\cdots$ in (\ref{vardef}) indicates the presence of terms with six
 or more  powers of fermion parameters that arise from
iterating the supersymmetry transformations.  The fermionic
contribution to the vector potential begins with a term that is
quartic in fermions
\be
\hat
A_{\mu} = {1\over 4!} \varepsilon_{ABCD}\zeta^{A} \sigma_{\mu\nu}
\zeta^{B} D^{\nu} ( F^{a}_{\lambda\kappa} \zeta^{C}
\sigma^{\lambda\kappa} \zeta^{D})  +\cdots \: ,
\label{ahdef}
\ee
where $\cdots$ indicates terms with eight  or more fermion
parameters.   The terms of higher order
in the fermions, which are not displayed in (\ref{vardef}) and
(\ref{ahdef}), involve not only the combination $\zeta$, but further
depend on the broken  conformal supersymmetry moduli
$\bar\xi$.\footnote{For example, there is an extra term of the form
$\varepsilon_{ABCD}\zeta^{A} \sigma_{\mu} \bar\xi^{B}  (
F^{a}_{\lambda\kappa} \zeta^{C} \sigma^{\lambda\kappa} \zeta^{D})$
in (\ref{ahdef}).}
The expression $\hat A_\mu$ contributes both to the self-dual field
strength and the anti self-dual field strength\footnote{We thank
S. Vandoren for pointing out that our statement in \cite{bgk} was
incorrect.}.

The calculation of the Wilson loop to leading order in $g_{_{YM}}$
involves substituting expressions (\ref{vardef}) and (\ref{ahdef})
into (\ref{wildef}).  The sixteen fermionic supermoduli integrals are
saturated, in principle, by expanding the exponential to extract the
terms with sixteen powers of fermionic coordinates.  This involves
dealing with complicated combinatorics that arises from various powers
of $\hat\varphi$'s and $\hat A$'s.  Significant cancellations between
the various terms should arise, just as there are in the calculations
of the instanton contribution to composite gauge invariant operators
\cite{bgkr,bkrs}. Nevertheless, explicit evaluation of the integral
appears to be prohibitively difficult and we will finesse it by making
extensive use of the (super)symmetries in a manner that generalizes
the purely bosonic case.

In section \ref{instsupsp} we will consider the extension of the
six-dimensional representation of the bosonic model to the
superconformal setting by introducing four four-component  Grassmann
spinors, $\Theta^A$, that are chiral spinors of $SO(4,2)$.  These
fermions will be associated with the Grassmann coordinates of a
supercoset that parameterizes the supersymmetries that are broken by
an instanton.  By means of  a more or less standard construction we
will obtain a representation of the $SU(2,2|4)$ superalgebra and its
$OSp(2,2|4)$ subgroup  in  terms of bosonic and fermionic coordinates
belonging to this supercoset.    In this way  we will represent the
supergroups relevant for the Wilson loop calculation in a chiral
fashion suitable for instanton computations.  Some of the details of
this supercoset construction are contained in appendix \ref{supvecs}.

The integration over the supermoduli, $(x_0,\rho_0, \eta,\bar\xi)$
will be considered in section \ref{N4oneinst}.  Once again we will
move the instanton to the centre of the loop by making use of the
residual $OSp(2,2|4)$.  This means moving it to the point $x_0 =0$ and
$\eta = \bar\xi =0$. This again allows the integration over the
supermoduli to be carried out as if the theory were abelian.  In this
manner we end up with an expression for the Wilson loop density on the
supermoduli space.  The form of this density apparently allows the
fermionic variables to be eliminated by a change of bosonic
integration variables, so that the integral over the Grassmann
variables formally vanishes.  However, this neglects the fact that the
bosonic integral diverges near the boundary of moduli space (the
$AdS_5$ boundary) and has to be regulated. Any regulator necessarily
introduces a dependence on the fermions at the boundary.

An ideal regulator would respect the Poincar\'e symmetries described
in section \ref{expect}.  This would require a detailed analysis of
the instanton contributions in the theory with $SU(3) \to SU(2) \times
U(1)$ in the limit of large symmetry breaking, which we have not
carried out.   Instead, in section 5 we shall simply cut off the
integration over the moduli in a manner that does not respect the
Poincar\'e supersymmetries.  We can anticipate that breaking supersymmetry
in such a manner will lead to a spurious
dependence on the cutoff that will have to be subtracted
to restore superconformal invariance.  After a certain amount of work
the fermionic integrations will be performed explicitly, leading to a
density on the bosonic moduli space.  This has the form of a
complicated tensor that contracts generators in the coset
$SU(2,2)/SO(2,2)$ (or $SO(5,1)/SO(3)\times SO(2,1$) in the euclidean
theory) acting on the bosonic loop.  The bosonic integration has the
form of the integral of a total divergence so, using Gauss' law, the
result is given by a boundary contribution, as anticipated.
Performing the generalized angular momentum algebra with the aid of
the algebraic software package REDUCE enables the explicit calculation
of   the Wilson loop expectation value.  As expected, the bulk
divergences of the bosonic theory do not arise in the supersymmetric
case.  However, the result of our calculation does have a linear
divergence proportional to the perimeter of the loop\footnote{In the 
preliminary description of our calculation \cite{bgk} we incorrectly 
assumed that this divergence would be absent.}.

As argued above, this perimeter divergence is a non
conformally-invariant artifact of our cutoff procedure.  It can be
eliminated by absorbing it into a counterterm for the mass
parameter of the test particle that defines the loop.  This leaves a
finite result for the loop expectation value, which is consistent with
conformal symmetry.  Furthermore, we will see in  section
\ref{subsubtwo} that the expression for the straight Wilson line has a
pure linear divergence of the same value per unit length as for the
circular loop, but without the subleading  finite part.  Therefore, the same
mass counterterm eliminates the divergence and
leads to the vanishing of the instanton contribution to the straight
line expectation value as required by the supersymmetry argument in
section \ref{expect}.

In the concluding section we will discuss the generalization of these
calculations to the gauge group $SU(N)$ in the semi-classical limit,
which is the limit of weak 't Hooft coupling when $N\to \infty$.  It
would be good to be able to say something about the limit of strong 't
Hooft coupling, which is of relevance for comparison with the
supergravity description but this is beyond explicit calculation.  We
will, however, make some comments on the way in which the instanton
contributions to the Wilson loop might be reconciled with the
$SL(2,\bbZ)$ Montonen--Olive duality of the $\N4$ theory and its image in
type IIB string theory in $AdS_5\times S^5$. The arguments given will
be qualitative.  We will also describe the calculation of the
instanton contribution to the correlation functions of the Wilson loop
with  gauge invariant composite operators in.  In fact, these
calculations are often easier than the pure Wilson loop calculation
and result in manifestly finite expressions.  In particular, we will
give a rather persuasive and simple argument that $d \langle W\rangle
/d\vartheta \neq 0$, as follows if $\langle W \rangle$ has an
instanton contribution.


\section{Symmetries of the circular loop}
\label{symmetries}

The ${\cal N}=4$ supersymmetric Yang--Mills theory  has a
superconformally invariant phase in which the scalar field expectation
values are zero.  The infinitesimal generators of the four-dimensional
conformal group, $( P_{\mu}, J_{\mu\nu}, D, K_{\mu})$, with conjugate
parameters $(a^\mu, \omega^{\mu\nu}, \lambda, b^\mu)$, have the
following action on the space-time coordinates
\be
\delta x^{\mu} = a^{\mu} + \omega^{\mu\nu} x_{\nu} + \lambda
x^{\mu} - x^{2} b^{\mu} + 2 b\cdot x x^{\mu} \: ,
\label{confdef}
\ee
where $x^2 \equiv \eta_{\mu\nu}x^\mu x^\nu$ and $\eta_{\mu\nu} = {\rm
diag}\, (+---)$.  The fifteen transformations of the $SU(4) \approx
SO(6)$ R-symmetry  group (with parameters  $\omega^{ij}$) have the
form
\be
\delta y^{i} = \omega^{ij} y_{j} \: .
\label{sufourdef}
\ee
 In addition, there are  four Poincar\'e
supersymmetries with generators $Q_{A}^{\alpha}$ and
$\bar{Q}^{A}_{\dot\alpha}$, as well as four superconformal symmetries
with generators $S^{A\alpha}$ and  $\bar{S}_{A\dot\alpha}$.  These
generators form the algebra associated with the supergroup
$SU(2,2|4)$, in which the fermionic generators satisfy the relations
\bea
\left[P^\mu,S_{\alpha}^A \right] &=&
\sigma_{\alpha\dot\alpha}^\mu\bar{Q}^{\dot\alpha A}  ,
\qquad \quad [K^\mu, Q_{\alpha A}] =
\sigma_{\alpha\dot\alpha}^\mu \bar{S}^{\dot\alpha}_A
\: , \nn\\
\left[D, Q_{\alpha A}\right] &=& -{1\over 2} Q_{\alpha A}
\: ,\qquad\quad\ \ \ \,
{}[D, S_{\alpha}^A] = +{1\over 2} S_{\alpha}^A\: ,\nn \\
\{Q_{\alpha A},\bar{Q}_{\dot\alpha}^B \}
&=& 2 \delta_A^B\sigma_{\alpha\dot\alpha}^\mu P_\mu
\: ,\qquad
\{S_{\alpha}^A ,\bar S_{\dot\alpha B} \}
= 2 \delta_B^A\sigma_{\alpha\dot\alpha}^\mu K_\mu
\: , \nn\\
\{\bar Q_{\dot\alpha}^A, S_{\alpha}^B\} &=& 0
\: , \qquad
\{\bar Q_{\dot\alpha}^A, Q_{\alpha\,B}\} = 0\: , \qquad
\{\bar S^{\dot\alpha}_A, S_{\alpha}^B\} =0\: , \nn\\
\{Q_{\alpha A},  S_{\beta}^B \} &=&
{1\over 2} \delta_A{}^B (\sigma^{\mu\nu})_\alpha^\beta J_{\mu\nu} +
2 \delta_A^B \delta_{\alpha}^\beta D \ + 2 \delta_{\alpha}^\beta T_A^B
\: .
\label{supconalg}
\eea
In addition, $Q$, $\bar Q$, $S$ and $\bar S$ transform as $SO(3,1)$
spinors of the appropriate chirality and as ${\bf 4}$'s or ${\bf{\bar
4}}$'s of $SU(4)$.  There is also a central $U(1)$ generator that acts
trivially on the  elementary fields and local composite operators
formed from them.

We are now interested in determining the subgroup of  $SU(2,2|4)$ that
leaves the circular loop, defined by
\be
(x^1)^{2} + (x^2)^{2} = R^{2},\qquad
x^3(s) = 0 , \qquad
x^0(s) = 0, \qquad
\dot{y}^{i}(s) = |\dot{x}| n^{i} \: ,
\label{circloop}
\ee
invariant up to reparametrizations.  Since we have fixed a direction
($n^{i}$) in the internal space, the loop is only invariant under  an
$SO(5) \approx Sp(4)$ subgroup of the $SU(4) \approx SO(6)$ R-symmetry
group.  It is convenient to define the $Sp(4)$ singlet
\be
\Omega_{AB} = n_{i} {{\widehat\Gamma}}^{i}_{AB}\: .
\label{spsiong}
\ee
The coordinates in the plane of the loop will be denoted by $(x^{l}) =
(x,y) \equiv (x^{1}, x^{2}) $ and those transverse  to the plane by
$(x^{t}) = (z,t) \equiv  (x^{3}, x^{0})$.  We will also define the
quantities $x_l^2 \equiv - x^l\, x_l$  and $x_t^2\equiv - x^t\, x_t$
so that $x_l^2\ge 0$ and (in the Wick-rotated theory)  $x_t^2\ge 0$.  The
action of   infinitesimal $SO(4,2)$ transformations on these
coordinates is
\bea
\delta x^{l} &=& a^{l} + \omega^{lm} x_{m} +
\omega^{lt}x_{t} + \lambda x^{l} + (x_l^{2} + x_t^{2}) b^{l} +
2 (b^{m}x_{m} + b^{t} x_{t} ) x^{l}\nn \\
\delta x^{t} &=& a^{t} +
\omega^{tm} x_{m} + \omega^{ts}x_{s} + \lambda x^{t} +(x_l^{2} +
x_t^{2}) b^{t} + 2 (b^{l}x_{l} + b^{s} x_{s} ) x^{t} \: .
\label{confac}
\eea
We are looking for the subset of these transformations that
preserves the loop at $x^2_{l} = R^{2}$, $x^{t} = 0$.  Along
the loop the transformations in (\ref{confac}) become
\bea
\delta x^{l} &=& a^{l} + \omega^{lm} x_{m} + \lambda x^{l} +
R^{2} b^{l} + 2 b^{m}x_{m} x^{l} \nn \\
\delta x^{t} &=& a^{t} + \omega^{tl} x_{l} + R^{2} b^{t} \: .
\label{prescon}
\eea
The condition $\delta x^t=0$ implies
\be
a^{t} = - R^{2} b^{t}\: , \qquad \omega^{tl} = 0 \: .
\label{trancon}
\ee
The condition $x_l \delta x^l =0$ is imposed along the loop by
contracting the first equation in (\ref{prescon}) with $x_{l}$,
giving
\be
0 = a^{l} x_{l} - \lambda
R^{2} - b^{l}x_{l} R^{2} \: ,
\label{looptwo}
\ee
which implies
\be
a^{l} = R^{2} b^{l}\: , \qquad \lambda = 0 \: .
\label{loopag}
\ee
Both $\omega^{lm}$ and $\omega^{ts}$ remain undetermined. The
resulting invariance of the loop is generated by  the six generators,
$J_{xy}$, $R^{2} P_{x} + K_{x}$, $R^{2} P_{y} + K_{y}$, $J_{zt}$,
$R^{2} P_{z} - K_{z}$ and  $R^{2} P_{t} - K_{t}$. In addition to the
obvious rotations in the $(x,y)$ plane and boosts in the $(z,t)$
plane, there are four combinations of translations and conformal
boosts.  These combinations  will be denoted by
\be
\Pi^{+}_{l} = R P_{l} + {1\over R} K_{l}\: ,
\qquad \Pi^{-}_{t} = R P_{t} - {1\over R} K_{t} \: .
\label{piplusmin}
\ee
These generators define the algebra $SO(2,2) = SO(2,1)\times
SO(2,1)$,
\bea
{}[ \Pi^{+}_{x}, \Pi^{+}_{y} ] &=& J_{xy},
\qquad [J_{xy}, \Pi^+_y] = \Pi^+_x, \qquad
[J_{xy}, \Pi^+_x] = - \Pi^+_y,
\nn \\
{}[ \Pi^{-}_{z}, \Pi^{-}_{t} ] &=& J_{zt}  ,
\qquad [J_{zt}, \Pi^-_t] = \Pi^-_z, \qquad
[J_{zt}, \Pi^-_t] =   \Pi^-_z\: ,
\label{otwoalg}
\eea
which is a subalgebra of $SO(4,2)$.  Note that the generators
$\Pi^-_l$ and $\Pi^+_t$ are in the coset $SO(4,2)/SO(2,2)$.  The ten
$Sp(4)$ generators that leave the loop invariant are given by the
symmetric combinations of the $SU(4)$ generators
\be
T_{AB} = T_{A}{}^C \Omega_{CB} + T_{B}{}^C \Omega_{CA} \: .
\label{spfour}
\ee
The symplectic metric $\Omega_{AB}$ and its inverse are used to lower
and raise the indices $A, B, \dots$.

We now want to find the fermionic part of the superconformal group
that leaves the loop invariant.  Since the  bosonic symmetry that
preserves the loop, $SO(2,2)\times Sp(4)$,  is the bosonic part of the
supergroup $OSp(2,2|4)$ (which is a subgroup of $SU(2,2|4)$)  this is
a natural candidate for the invariance group of the loop.  To verify
that this is indeed the case we want to first identify the  Killing
spinors  that satisfy
\be
\dot{x}_{\mu} \sigma^{\mu} \bar\kappa_{A}(x)
= \dot{y}_{i} {{\widehat\Gamma}}^{i}_{AB} \kappa^{B}(x) \: ,
\label{kiljoy}
\ee
where $\kappa^A = \eta_{_{\oplus}}^A + x\cdot \sigma
\bar\xi_{_{\oplus}}^A$ and
$\bar\kappa_A = \xi_{_{\oplus}\, A} + x\cdot \bar\sigma
\bar\eta_{_{\oplus}\, A}$ with $\eta_{_{\oplus}\, \alpha}^A$,
$\bar\xi_{_{\oplus} \dot\alpha}^A$, $\bar\eta_{_{\oplus}\, A}$ and
$\xi_{_{\oplus} \,A}$ constant complex  2-component
spinors.  Using the parametrization $\dot{x}^{l} = \omega
\varepsilon^{lm} x_{m}$ and $\dot{y}_{i} {{\widehat\Gamma}}^{i}_{AB} =
\omega R \Omega_{AB}$ and taking into account that $x^t = 0$ along the
loop gives
\be
\varepsilon^{lm} x_{m} \sigma_{l} (\bar\eta_{_{\oplus}\, A} + x_{n}
\bar\sigma^{n} \xi_{_{\oplus}A}) = R \Omega_{AB}
( \eta_{_{\oplus}}^{B} + x_{m} \sigma^{m} \bar\xi_{_{\oplus}}^{B})
\: .
\label{implicit}
\ee
More explicitly,
\be
(x_1 \sigma_2 - x_2 \sigma_1) [\bar\eta_{_{\oplus}\, A} + (x_1
\sigma_1 + x_2 \sigma_2) \xi_{_{\oplus}\, A}] =
R \Omega_{AB} [ \eta_{_{\oplus}}^{B} + (x_1
\sigma_1 + x_2 \sigma_2)\bar\xi_{_{\oplus}}^{B}] \: .
\label{moreso}
\ee
Using $x_1^2 + x_2^2 = R^2$,  this implies
\be
(x_1 \sigma_2 - x_2 \sigma_1) \bar\eta_{_{\oplus}\, A} + R^2 \sigma_2
\sigma_1
\xi_{_{\oplus}\,A} = R \Omega_{AB} [ \eta_{_{\oplus}}^{B} + (x_1
\sigma_1 + x_2
\sigma_2)\bar\xi_{_{\oplus}}^{B}] \: ,
\label{xone}
\ee
so that
\be
\xi_{_{\oplus}\, A} = {1\over R} \Omega_{AB} \sigma^{12}
\eta_{_{\oplus}}^{B} \: , \qquad
\bar \xi_{_{\oplus}}^{A} = {1\over R} \Omega^{AB} \bar\sigma^{12}
\bar\eta_{{_\oplus}B} \: .
\label{xidef}
\ee
The resulting fermionic symmetries are thus generated by
\be
G_{A} = \sqrt{R}\sigma^{12} Q_{A} + {1\over \sqrt{R}}\Omega_{AB}
S^{B}
\: , \qquad
\bar{G}^{A} = \sqrt{R}\bar\sigma^{12} \bar{Q}^{A} +
{1\over \sqrt{R}}\Omega^{AB} \bar{S}_{B} \: .
\label{gdeff}
\ee

The anti-commutators of the supersymmetry generators are
\bea
\{ G_A, \bar{G}_B \} &=& R \sigma^\mu P_\mu + \sigma^{12} \sigma^\mu
\sigma^{12} {1\over R} K_\mu =
{1\over R} \sigma^l \Pi^+_l + {1\over R} \sigma^t \Pi^-_t \nn \\
\{ G_A, G_B \} &=&  \Omega_{AB}
(\sigma^{12} \sigma^{\mu\nu} \sigma^{12} +  \sigma^{\mu\nu})
J_{\mu\nu} +
{1\over R^2} (T_{AB} + T_{BA}) \nn \\
&=& 2 \Omega_{AB}
(\sigma^{xy} J_{xy} +  \sigma^{zt} J_{zt}) +
(T_{AB} + T_{BA}) \: .
\label{anticom}
\eea
In a more compact notation the surviving supersymmetry generators
$G$ and $\bar{G}$ can be packaged into $G_A^a$, where $a$ is an index
of the ${\bf (2,2)}$ of $SO(2,2)$, in which case the supersymmetry
algebra reads
\be
\{ G^a_A, G^b_B \} = \Omega_{AB} J^{ab} + T_{(AB)} \: ,
\label{compsus}
\ee
where the six generators of $SO(2,2)$,
$( \Pi^+_l, J_{xy},\Pi^-_t, J_{zt})$ have been assembled into
$J^{ab}$.
The remaining  commutation relations of the $OSp(2,2|4)$ algebra are
\bea
{} [J_{ab}, J_{cd}] &=& H_{bc} J_{ad} + {\rm perms}\: ,\qquad
{} [T_{AB}, T_{CD}] = \Omega_{BC} T_{AD} + {\rm perms}\: ,\qquad
{} [T_{AB}, J_{ab}] =0
\nn \\
{} [J_{ab}, G_{Ac}] &=& H_{bc} G_{Aa} - H_{ac} G_{Ab}\: ,\qquad
{} [T_{AB}, G_{Ca}] = \Omega_{BC} G_{Ac} +  \Omega_{AC} G_{Bc} \: ,
\label{smallalg}
\eea
where $H$ (to be read as `capital $\eta$') denotes the $SO(2,2)$
invariant metric tensor.


\section{One-instanton contribution in the bosonic model}
\label{pureYM}

Before tackling the complete integral over the bosonic and
fermionic instanton moduli in the $\calN =4$ theory,
we will consider some essential features
that  arise purely from the bosonic integrations.  The expression for
the Wilson loop that is obtained
by simply substituting the BPST instanton solution into
(\ref{wildef}), setting the fermionic variables to zero and ignoring
the fermionic integrations will be referred to as the `bosonic model'
(it is {\it not} the Wilson loop of pure Yang--Mills, which has a different
and non conformally-invariant  measure).

In calculating the Wilson loop we will make use of the fact that the
form of the instanton profile is invariant under euclidean conformal
transformations with the understanding that the  moduli are
transformed by compensating conformal transformations. In particular,
it is possible to transform a one-instanton configuration into an
equivalent one by acting with an element of $SO(3)\times SO(2,1)$,
which maps the loop onto itself.  Since the moduli space is the
five-dimensional anti de-Sitter space spanned by $(x_0^\mu, \rho_0)$,
we will need  to represent the action of $SO(5,1)$ and $SO(3)\times
SO(2,1)$ on these coordinates.  These groups act nonlinearly on
euclidean $AdS_5$ so it is convenient to represent them in terms of a
six-dimensional space with one time-like coordinate which has
signature $(5,1)$ on which the groups act linearly.  This is the
procedure first introduced by Dirac \cite{dirac}  in order to describe
the action of $SO(4,2)$ on four-dimensional Minkowski space.

\subsection{Six-dimensional representation of $SO(4,2)$
and $SO(2,2)$}
\label{dirac6dim}

The flat  six-dimensional coordinates $X_M$ (where $X_0$ and $X_4$ are
time-like) are taken to satisfy the rotationally  invariant constraint
\be
X^2 \equiv \eta_{MN} X^M X^N \equiv
\eta_{\mu\nu} X^\mu X^\nu + (X_4)^2  - (X_5)^2  = \ell^2,
\label{sixcon}
\ee
where $\ell$ is a constant scale and for the moment we are using
six-dimensional metric $\eta_{MN} = {\rm diag}(+---+-)$ appropriate to
Minkowski signature in four dimensions $M=0,1,2,3$.  The euclidean
case is obtained by the Wick rotation $X_0 \to i X_0$.  The constraint
is solved in terms of five-dimensional coordinates that parameterize
$AdS_5$ with scale $\ell$.  A conventional parameterization of $AdS_5$
in terms of $x^\mu$, $\rho$, is obtained by the identifications
\bea
X^\mu = \ell {x^\mu \over \rho}\: , \qquad
X^4 = {1\over 2} \left( \rho + {\ell^2 - x^2 \over \rho}\right)\: ,
\qquad
X^5 = {1\over 2} \left( \rho - {\ell^2 + x^2 \over \rho}\right) \: ,
\label{solvec}
\eea
which represents an $AdS_5$ hypersurface in $\bbR^6$.  Inverting
these conditions gives
\be
x^\mu  =  \ell {X^\mu \over X^4 - X^5} \: , \qquad
\rho  =  {\ell^2 \over X^4 - X^5} \: , \qquad \rho^2 - x^2  =  \ell^2
{X^4 + X^5 \over X^4 - X^5} \: .
\label{invconx}
\ee
Recall that in the above parametrization the boundary of $AdS_5$ is at
$\rho=0$. In the following we will also find it useful to define
$X^{\pm}=X^4\pm X^5$.

It is then easy to check that the Lorentz transformations
on the six-dimensional coordinates, generated by
\be
L_{MN} = X_M
\partial_N - X_N \partial_M \: ,
\label{sixtrans}
\ee
induce $SO(4,2)$ transformations on $AdS_5$, with the identifications
of the fifteen generators
\be
J_{\mu\nu} = L_{\mu\nu}
\, , \qquad D = L_{45} \, , \qquad \ell P_\mu = L_{4\mu} + L_{5\mu}
\, ,
\qquad K_\mu = \ell (L_{4\mu} - L_{5\mu}) \: .
\label{fifpars}
\ee
Furthermore the trivial six-dimensional integration measure is
equivalent, after the constraint, to the $AdS_5$ measure
\be
\ell^{-4}\int \delta(X^2 -\ell^2) \,  d^6X =
\, \int {d^4 x d\rho\over \rho^5} \: .
\label{measeq}
\ee
The boundary is mapped into itself under the $SO(4,2)$ transformations
and  the familiar four-dimensional action of $SO(4,2)$ results from
the boundary  limit in which $\rho \to 0$ \cite{dirac}.  The
generators $\Pi^\pm_\mu$ introduced earlier are expressed in this
six-dimensional notation as
\be
\Pi^\pm_\mu = \left({\ell\over R} \pm {R\over \ell} \right)L_{4\mu} +
\left({\ell\over R} \mp{R\over \ell} \right)L_{5\mu}.
\label{sixpi}
\ee

Our aim is to consider a loop of radius $R$ and to identify $AdS_5$
with the one-instanton moduli space. For this purpose we will need to
consider the euclidean theory obtained by Wick rotation, which
involves insertion of judicious factors of $i$.  Since the scale
$\ell$ drops out of all physical quantities it is convenient to choose
$\ell=R$ for most of the following (conformal invariance further
implies that $\la W \ra$ is a constant, independent of $R$).  In this
case the expressions (\ref{sixpi})  for $\Pi^\pm_\mu$ become
particularly simple and the stability group generator $\Pi^+_l$ acts
only on $(X_L)= (X_4, X_l)$ while $\Pi^-_t$ acts only on
$(X_T)=(X_5,X_t)$.  We saw earlier that the loop at $\rho=0$, $|x_l|^2
= R^2$, $x^t=0$ is invariant (after a Wick rotation) under $SO(3)_L
\times SO(2,1)_T$ transformations.  These transformations are
described in the six-dimensional formalism as those that leave
invariant the quadratic form

\be
U = X_L^2 \equiv  (X_4)^2 - (X_l)^2
= {1\over 4} \left( \rho + {R^2 + x_{l}^2 + x_{t}^{2} \over
\rho}\right)^{2}
- R^{2}\, {x_l^{2} \over \rho^{2}} \: ,
\label{bosform}
\ee
where $X_L  = (X_1,X_2,X_4)$ are the components of $X_M$ that are
transformed by the $SO(3)_L$ subgroup while $X_T = (X_0, X_3, X_5)$
transform under the $SO(2,1)_T$ subgroup\footnote{For convenience we
are using the conventions
\bea
X_L^2 &=& (X_4)^2 - (X_1)^2 - (X_2)^2 \equiv (X_4)^2 - X_l^2\: ,
\nn\\
X_T^2 &\equiv& \mp (X_0)^2 + (X_3)^2 + (X_5)^2\equiv (X_5)^2 + X_t^2
\: ,
\label{convens}
\eea
with the $-$ sign for Minkowski signature and $+$ sign after the Wick
rotation   of the time coordinate.  This means that with
euclidean signature $X_t^2, X_T^2 \ge 0$ and therefore $X_L^2 = X_T^2+
R^2 \ge R^2$ after using the constraint.}.

For fixed $U$ equation (\ref{bosform}) defines a four-dimensional
hyperbolic surface in $AdS_5$, once the constraint (\ref{sixcon}) is
imposed.  Such four-surfaces of constant $U$ foliate the interior of
$AdS_5$ in such a manner that they all meet on a circle of radius $R$
centered on the point $x^\mu=0$ on the boundary at $\rho=0$.  In other
words,  all the four-surfaces are bounded by the Wilson loop if the
boundary of $AdS_5$ is identified with four-dimensional space-time
(see the figure).  The value of $U$ is in the range $R^2 \le U \le
\infty$.  There are two surfaces for every value of $U>R^2$, while the
surface with with the minimal value, $U=R^2$, is degenerate since it
is two-dimensional.  It is defined by $x_t=0$, $x_l^2 + \rho^2 =R^2$,
which is just the surface of minimal area embedded in $AdS_5$ which
bounds the loop of radius $R$ on the boundary and was considered in
\cite{mald}.

\FIGURE[!h]{
{\includegraphics[width=0.6\textwidth]{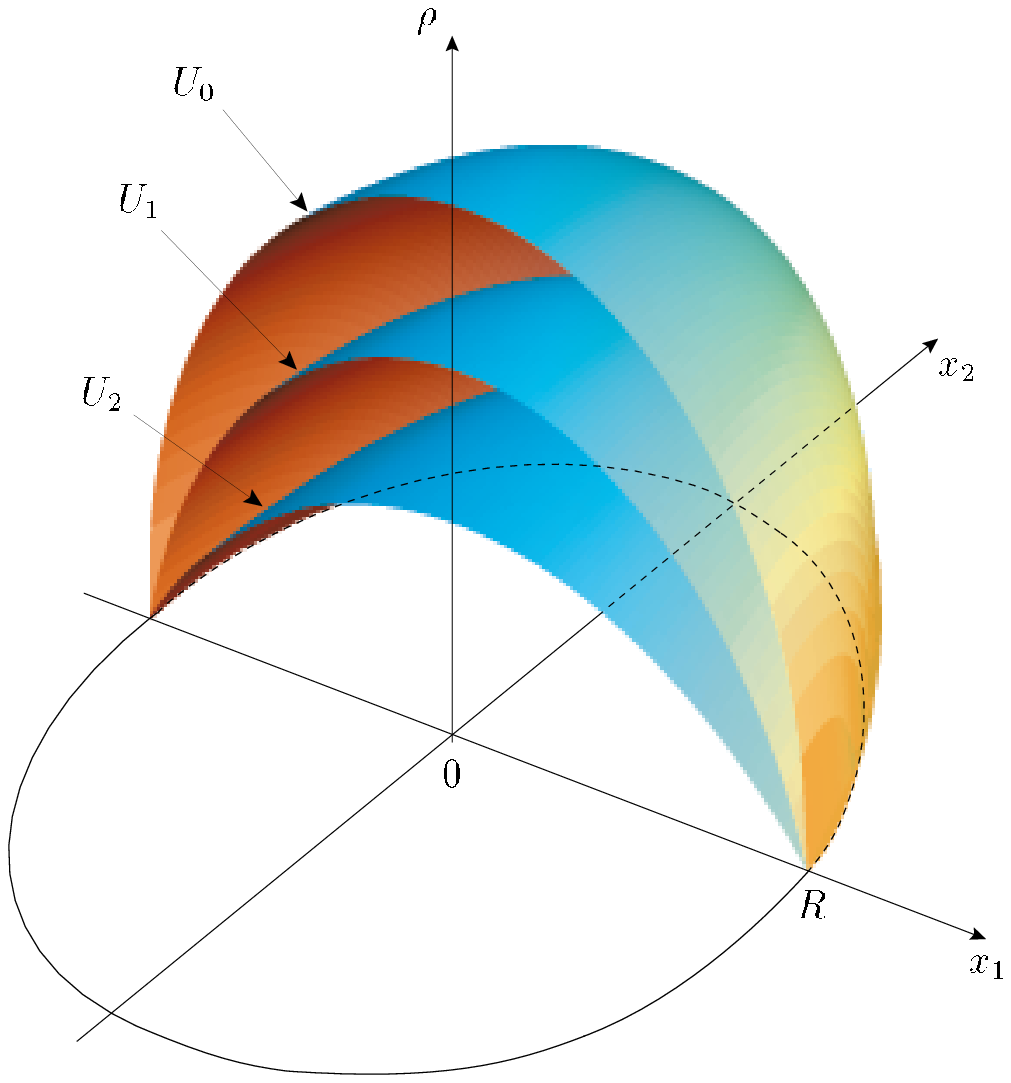}}
\caption{Surfaces of constant $U$ are $SO(2,2)$ orbits of codimension
$1$ in $AdS_5$.  The plot shows a section with $x_t=0$.
All surfaces with constant $U$ end on the loop $x_l^2 =R^2$, $x_t=0$
on the boundary, $\rho=0$. The surface with the minimal value,
$U=U_0= R^2$, coincides with the minimal
two-dimensional surface bounded by the loop.  The surfaces shown have
$U\ge R^2$ and intersect the $\rho$ axis with $\rho \le R$. The
surfaces that intersect the axis with $\rho > R$ have been omitted
from the figure.}
\label{surfig}
}


\subsection{The Wilson loop in the bosonic model}
\label{bosonic}

The Wilson loop expectation value in the toy bosonic
model is given by
\be
\la W_B \ra = \int {d^4 x_0 d\rho_0\over \rho_0^5}\,
 W_B[x(\cdot);x_0,\rho_0]\: ,
\label{bosfive}
\ee
where
\be
W_B[x(\cdot);x_0,\rho_0] = \frac{1}{2}\Tr \,\calP\,e^{i\int_\calC A\cdot \dot{x}}\: ,
\label{wbdef}
\ee
with $A$ denoting the standard BPST instanton solution,
and where we are temporarily adopting
 a notation that makes explicit that
$W_B$ is a functional of the points on a circle on the boundary of
$AdS_5$.

It will be important that $W_B$ is invariant under arbitrary $SO(3) \times SO(2,1)$
transformations that map the circle  into itself (in euclidean signature).
To show this, consider
the action of a general $SO(3)\times SO(2,1)$ transformation,
$x\to   \gamma(x) \equiv \tilde x $ on $W_B[x(\cdot);x_0,\rho_0]$.  On the one hand,
this transformation maps the circle
into itself so that, after a reparameterization of the circle (under which
$W_B$ is invariant), the points on the circle are not transformed and therefore
$W_B[x(\cdot); x_0,\rho_0] = W_B[\tilde x(\cdot);x_0,\rho_0]$.
On the other hand the transformation of $x$ in the instanton solution  is
equivalent (up to an irrelevant gauge transformation)
to a transformation on the instanton moduli
$(x_0, \rho_0) \to (\tilde x_0, \tilde \rho_0) = \gamma^{-1}(x_0,\rho_0)$.  Therefore
\be
W_B[x(\cdot); x_0,\rho_0] = W_B[x(\cdot); \tilde x_0, \tilde \rho_0] \: ,
\label{moveinst}
\ee
and so the density $W_B$ depends only on the choice of $SO(3) \times
SO(2,1)$ orbit, which can be labeled by
the value of the invariant, $U= X_L^2$, defined in the previous
subsection.

We will soon find it useful to choose an appropriate $SO(2,2)$ transformation, $\gamma$,
that moves the instanton to a point
$(\tilde x^\mu_0 =0, \tilde \rho_0)$, which is at the centre of the loop.
The scale
 of the transformed instanton is fixed by the invariance of $U=X_L^2
=X_L^2 (\tilde{x}_0 = 0; \tilde{\rho}_0)$, which implies
\be
{1\over 4} \left( \rho_0 + {R^2 + x_{l}^2 + x_{t}^{2}
\over \rho_0}\right)^{2} - R^{2}{x_{l}^{2} \over \rho_0^{2}}
={1\over 4}  \left({R^2\over \tilde{\rho}_0} +\tilde{\rho}_0\right)^2\: ,
\label{invcent}
\ee
or
\be
\tilde \rho_0 = |X_L| - |X_T|\: .
\label{solverho}
\ee
This expression relates the parameters of an instanton at a generic
position in moduli space to one at $\tilde{x}_0=0$ with a scale $\tilde
\rho_0(x^\mu,\rho)$.   The explicit transformation that moves the instanton from one
point to another along an $SO(3) \times SO(2,1)$ orbit is a group element of the form
$\exp(a^l\Pi^+_l  + a^t\Pi^-_t)$, where the parameters
$a^l(x_0^\mu,\rho_0)$ and $a^t(x_0^\mu,\rho_0)$ are specific functions
of the collective coordinates (which we do not need in the following).

It will prove efficient to express the integral over the instanton moduli space,
(\ref{bosfive}), as a six-dimensional integral with flat measure
together with a $\delta$-function constraint
\bea
\la W_B \ra &=& {1\over R^4}\, \int d^6 X_0\,
\delta( X_0^2 - R^2)\, \frac{1}{2} \Tr {\cal P}\, 
e^{i\int_\calC A\cdot \dot{x} ds}  \nn \\
&=& {1\over R^4}\, \int d^6 X_0 \, \delta( X_0^2 - R^2)\, W_B(X_0)\: ,
\label{bossix}
\eea
where
\be
W_B(X_0) = \frac{1}{2}\Tr P\,
\exp\left(i\int {\eta^a_{\mu\nu} (x^\mu - x^\mu_0) \sigma_a \over
\rho_0^2 + (x - x_0)^2}\, \dot x^\nu\, ds\right)\: .
\label{newwdef}
\ee
The standard one-instanton solution in a `nonsingular gauge' (a
gauge in which the singularity arises at $|x|=\infty$) has been
substituted for $A_\nu$ in (\ref{newwdef}).
The Pauli matrices $\sigma^a$ describe the
$SU(2)$ colour symmetry. Expression (\ref{bossix}) does not include
the correct prefactor that arises from gaussian fluctuations, which we
are ignoring in the  bosonic model.\footnote{Various overall numerical
constants will be dropped from the expressions for the Wilson loop but
they will be reinstated in the final result.}

In the special case in which the instanton is at the centre of the
loop the path ordered exponential simplifies since $\sigma_a
\eta^a_{\mu\nu} \, x^\mu \, \dot x^\nu ds= R^2 \sigma^3 d\phi$, where
$0\le \phi\le 2\pi$ is the angle around the loop, which has been taken
to lie in the $(x^1,x^2)$ plane.  So in this special configuration the
exponent is proportional to $\sigma^3$ and the path ordering becomes
trivial, as in the abelian theory.  More generally, it should always be
possible to choose a gauge in which the connection along a given curve
is a non-vanishing constant (analogous comments concerning  the
maximally abelian gauge appear in \cite{ht,Browera,Browerb}).  The
integration over the angle $\phi$ followed by evaluation of the trace
leads to (dropping the subscript 0)
\be
W_B(X)=   \cos\left({2\pi R^2
\over R^2 + \tilde{\rho}^2}\right) \: .
\label{cenres}
\ee
The value of $\tilde\rho$ may be expressed in terms of the
invariant  $X_T^2 = U - R^2$ by  using (\ref{solverho}), giving
\be
{2 \pi R^2 \over R^2 + \tilde{\rho}^2} = \pi + {\pi |X_T| \over
\sqrt{X_T^2 + R^2} } \: ,
\label{specip}
\ee
so the  Wilson loop
density can be expressed as
\be
W_B(X) =  - \cos\left( {\pi |X_T| \over \sqrt{X_T^2 + R^2} }
\right)\:.
\label{wbdeff}
\ee

Since $W_B$ depends only on one parameter $|X_T| =\sqrt{U - R^2}$ that
labels the $SO(3)\times SO(2,1)$ orbits,  the
five-dimensional integration over the bosonic moduli in (\ref{bossix}) reduces to
a one dimensional integral over $|X_T|$ with a measure that is proportional
to  the volume of the orbit labeled by $|X_T|$. This leads to an infinite value
due to the divergence
of the $SO(2,1)$ volume near the boundary of moduli space ($\rho =0$ in
our five-dimensional coordinates).  Similar
considerations will also apply to the $\calN=4$ theory and
we therefore anticipate the need
to introduce a regulator that suppresses point-like instantons.  Other regions
of moduli space  also lead to divergences in the bosonic model.

We will proceed by explicitly performing
the two-dimensional $X_l$ integration in (\ref{bossix}) to eliminate the
$\delta$ function and regulating the remaining integrations by introducing a large
$X_4$ cutoff,
\be
X_4\le \Lambda \equiv {R^2\over\epsilon},
\label{largecut}
\ee
where $\epsilon$ is a small scale with dimensions of length.  This
cutoff manifestly preserves a $SO(5)$ subgroup of $SO(5,1)$.  Performing the $X_4$
integral  gives
\bea
\la W_B \ra_\epsilon & = & - {2\pi\over R^4}\,  \int_{|X_T|\le
\sqrt{\Lambda^2 - R^2}}
d^3 X_T  \,
\left(\Lambda - \sqrt{X_T^2 + R^2}\right)\,
\cos \left( {\pi |X_T| \over \sqrt{X_T^2 + R^2}} \right)\nn\\
&=& -2\pi \int_{|{\bf X}| \le \sqrt{R^2/\epsilon^2 -1}} d^3{\bf X} \,
\left({R\over \epsilon} - \sqrt{|{\bf X}|^2 + 1}\right)\,
\cos \left( {\pi |{\bf X}|\over \sqrt{|{\bf X}|^2 + 1}} \right)
\: ,
\label{finre}
\eea
where ${\bf X} \equiv  X_T/R$.  This integral diverges
 when $\epsilon\to 0$.  The leading divergence  is of order
 $\epsilon^{-4}$ and arises from the
regions in which the instanton has a scale that is very much greater
 than $R$ and those in which it has a fixed scale but is
very far from the loop.  This bulk divergence has no analogue in the $\calN =4$ case
to be considered later.  The region where the instanton scale is very much smaller than $R$
leads to the divergence of order $\epsilon^{-1}$ associated with the volume of the
$SO(3) \times SO(2,1)$ orbits and which will remain as an important issue in the analysis
of the supersymmetric theory.


\subsection{Comments on the straight Wilson  line}
\label{lineambiguity}

If the curve ${\calC}$ is taken to be a straight line  there is room
for potential confusion.   One well-defined way to obtain a straight
Wilson line is  to consider it to be a stereographic projection of a
circular loop passing through the north pole of $S^4$ that is the
boundary of euclidean $AdS_5$.  Since the north pole is not a special
point on the sphere, in a conformally invariant theory this  gives a
result that is the same as that for a generic circular loop.  However,
this differs from the natural definition of a  straight Wilson line
defined directly in $\bbR^4$.  The latter corresponds to the starting
point of \cite{dg} where it was emphasized that the presence of a
conformal anomaly leads to a different expression  from the circular
loop\footnote{One way to think of this straight Wilson line is to
consider it to be the limit of a `thermal Wilson loop', or a Polyakov
loop, defined in $\bbR^3 \times S^1$, where the circular dimension has
infinite radius.}.  This will also be true in  the presence of an
instanton.

One subtlety in the analysis of the instanton contribution to the
straight line concerns the gauge independence of the calculation.  The
BPST instanton solution for the gauge potential in the so-called
`non-singular' gauge is, in fact, singular at the point at infinity.
Since this point coincides with a point on the straight line it is
best to avoid this gauge and use a `singular' gauge, in which the
gauge potential is singular at the point ${\bf x}={\bf x_0}$, $t=t_0$.
This gives the expression
\be
\langle W_B({\rm line})\rangle =
\int {d^3{\bf x}_0 dt_0 d\rho_0\over \rho_0^5}\, W_B^{s},
\label{srtrint}
\ee
for the expectation value of the Wilson line at $x^i =0$, where
\be W_B^{s} = {1\over 2} \Tr \calP
\left( \exp{ i \int_{-\infty}^{+\infty}
{\rho_0^2 \, {\bf{x}}_0 \cdot {\bf{\sigma}} dt \over
[\rho_0^2 + |{\bf{x}}_0|^2 + (t-t_0)^2]
[ |{\bf{x}}_0|^2 + (t-t_0)^2] }}\right)\: .
\label{singloop}
\ee
We have used the fact that the vector potential is proportional to
$\eta^a_{0i}(x^i - x_0^i)\sigma_a$ where $\eta_{01}^a = \delta_i^a$.
The $SU(2)$ connection in
this expression is  abelian since it always points along ${\bf{x}}_0
\cdot {\bf{\sigma}}$ so the path ordering is immaterial and the
integral is then easily evaluated giving
\be
\int_{-\infty}^{+\infty}
{\rho_0^2 dt \over
[\rho_0^2 + |{\bf{x}}_0|^2 + (t-t_0)^2]
[ |{\bf{x}}_0|^2 + (t-t_0)^2]} = {\pi\over|{\bf{x}}_0|} -
{\pi\over\sqrt{|{\bf{x}}_0|^2 + \rho_0^2}}\: ,
\label{singin}
\ee
so that
\bea
W_B^{s} &=& {1\over 2} \Tr \exp{\left({i \pi{\bf{x}}_0
\cdot {\bf{\sigma}}\over|{\bf{x}}_0|}
\left[ 1 - {|{\bf{x}}_0| \over\sqrt{|{\bf{x}}_0|^2 +
\rho_0^2}}\right]
\right)}
= -
\cos{\left( {\pi |{\bf{x}}_0| \over \sqrt{ \rho_0^2 +
|{\bf{x}}_0|^2}}
\right)}   \: .
\label{wilsing}
\eea
Starting, instead,  with the instanton in the `non-singular' gauge gives
a result for the Wilson loop expectation value with the opposite
sign.  However in that case there is a
subtlety because the gauge potential is actually
singular at infinity, which is a point on the
straight line.  To avoid this problem,
in the  following we shall choose the connection
in the `singular' gauge although the final results should not depend on this
choice.

The expression (\ref{wilsing}) has the same structure as
(\ref{wbdeff}), which refers to circular loops, but with $|X_T|/R$
replaced by $|{\bf{x}}_0|/\rho_0$.  However, this is a rather formal
correspondence since in both cases integration over the moduli gives a
divergent result.  In order to understand the connection between the
straight line and the circular Wilson loops more quantitatively we
will discuss the group theoretical relation between them.  In
euclidean signature the stability group of the straight line and the
circle are isomorphic subgroups of $SO(5,1)$ -- they are both $SO(3)
\times SO(2,1)$.  One of these subgroups is obtained from the other by
conjugation with an infinite boost generated by $\Pi^{-}_{l}$.
Alternatively, the straight line may be obtained from the circle by
inversion with respect to a point on the circle.   In the case of the
straight line the $SO(3)$ factor refers to rotations in the three
dimensions orthogonal to the line and the $SO(2,1)$ subgroup
corresponds to a dilation combined with a translation and a conformal
boost along the direction of the line. In the supersymmetric case the
surviving R-symmetry is $Sp(4)\approx SO(5)$, as for the circle.
There is, however, a crucial difference between the straight line and
the circle when a regulator is introduced.  In terms of the $AdS_5$
description of the regulated field theory the natural regulator for
the straight line defined in $\bbR^4$  is the Poincar\'e invariant
condition $\rho_0 \ge \epsilon$ which preserves the isometries of the
loop.  In this case the expectation value of the Wilson line 
has a finite contribution per unit
length, but diverges due to the integration over the length.  Upon
mapping the infinite line to a circle through the north pole of the
$S^4$ boundary of $B^5$ this cutoff prescription is pathological since
the $\rho=\epsilon$ surface touches the boundary at the north pole.
It is therefore not a good regulator for loops on $S^4$.  An
appropriate regulator in this case is obtained by cutting off the
moduli integrations on a spherical shell inside the $B^5$ boundary in
a manner that preserves $SO(5)$. As we will see, in the superconformal
case this distinction becomes of paramount importance since
supersymmetry implies that the  expectation value is  unity in
the case of the straight Wilson line while it has nontrivial dependence on
the coupling  for a circular loop.

The expression for the straight Wilson line in the presence of a
$\rho\ge \epsilon$ cutoff has the form
\bea
\la W_B({\rm line})\ra_\epsilon &=&
\int dt_0 \int d^3{\bf x}_0  \int_\epsilon^\infty {d\rho_0\over
\rho_0^5}\, W_B^{s}
=  - \int dt_0 \int_\epsilon^\infty {d\rho_0\over \rho_0^2}\int d^3
{\bf X}
\, \cos\left({\pi |{\bf X}|\over \sqrt{|{\bf X}|^2 + 1}}\right)\nn\\
&=&  - \int dt_0\,  {1\over \epsilon} \int d^3 {\bf X}
\, \cos\left({\pi |{\bf X}|\over \sqrt{|{\bf X}|^2 + 1}}\right),
\label{singint}
\eea
where ${\bf X} ={\bf x}_0/\rho_0$.  This is the integral of a constant
density over the infinite length of the line.  In the limit $\epsilon
\to 0$ the divergence in the density  is {\it identical} to that
arising from the $1/ \epsilon$ term in the last equation in
(\ref{finre}) that describes a circular loop of length $2\pi R$. As remarked
earlier, the
three-dimensional ${\bf X}$ integration gives rise to the bulk divergences
of the bosonic model which are not present in the supersymmetric theory.


\section{Instanton superspace and the $\calN=4$ Wilson loop}
\label{instsupsp}

The bosonic toy model is not a realistic
description of any bosonic quantum field theory.  In the case of pure
Yang--Mills the quantum measure of integration involves a ratio of 
determinants which is not scale invariant.  A detailed
computation of these determinants was performed in 't Hooft's seminal
paper on instanton calculus \cite{tHooft}.
In supersymmetric theories the fluctuation determinant
leads to  a dependence on the scale parameter $\mu$ of the form
\be
(\mu\rho)^{n_{B}- {1\over 2} n_{F}}
\label{betfun}
\ee
where $n_{B}$ and $n_{F}$ are the numbers of bosonic and fermionic
zero modes respectively. The exponent $n_{B}- {1\over 2} n_{F} = k
\beta_{1}$ in (\ref{betfun}) is the coefficient of the
$\beta$-function of $g_{_{YM}}$ at one-loop times the instanton
number.  Indeed for a generic ${\cal N} = 1$ supersymmetric theory
\be
\beta_{1} = \left(
{11\over 3} - {1\over 2}\times {4\over 3}\right) C_{A} - \sum_{R}
\left({1\over 2}\times  {4\over 3} + {2\over 6}\right) C_{R}\: ,
\label{beta}
\ee
where the $C_A$ term comes from the vector supermultiplet and the
$C_R$ terms comes from a sum of the chiral supermultiplets in
representations labelled $R$ ($C_A$ and $C_R$ being the appropriately
normalized Dynkin indices).  Index theorems imply that
\be
n_{B} = 4 k C_{A} \: , \qquad n_{F} = 2 k
C_{A} + 2 k \sum_{R} C_{R}\: .
\label{indth}
\ee
For (super-)conformal theories, such as ${\cal N}=4$ Yang--Mills,
$\beta = 0$ and the quantum measure for the collective
(super-)coordinates exactly coincides with the classical one. For
supersymmetric non superconformal theories the quantum measure is
deformed so it is `squashed'. But, in principle, one can still keep
track of the transformations under the sequence of steps that bring
the instanton to the centre of the circular loop.  Although this is
not the subject of the current paper it should be possible to
generalize the manipulations of the superconformal $\N4$ theory to
these cases.

In order to extend the analysis of section \ref{pureYM}
to ${\cal N} = 4$ supersymmetric Yang--Mills theory  we need to
include the sixteen fermionic collective coordinates, $\eta^A_\a$ and
$\bar \xi^A_\adot$. This means that we need to consider the extension
of the six-dimensional representations of $SO(4,2)$ and $SO(2,2)$ (and
their euclidean continuations) to the supersymmetric case.

\subsection{Six-dimensional chiral representation of $SU(2,2|4)$
and $OSp(2,2|4)$}
\label{super6dim}

Using the methods of \cite{supcos} we
will present a supercoset construction to represent the action of the
superconformal group   $SU(2,2|4)$ in terms of the six bosonic
coordinates $X^M$ and of four Grassmann spinors,
$\Theta^A_a$, where $a$ is a four-component spinor index appropriate
to a Weyl spinor in $D=6$ with signature $(4,2)$. This provides a
chiral representation of the one-instanton superspace.

The instanton solution breaks a subset of the bosonic and fermionic
symmetries in $SU(2,2|4)$. It is invariant under rotations modulo
gauge transformations and under
the linear combination $P_\mu + K_\mu/\rho^2$ of translational and 
conformal boost symmetries.
Altogether this means
that an  $SO(4,1)$ subgroup of $SO(4,2)$ is left unbroken. In
addition, the instanton preserves the $SU(4)_{R}$ symmetry.  The
supersymmetries that remain unbroken by the instanton are $\bar
Q^A_{\dot \alpha}$ and $S^A_\alpha$.   Putting these together means
that the $\N4$ instanton superspace may be described by the coset
\be
G/ H \equiv {SU(2,2|4) / Span \{ SO(4,1)\times SU(4);
\bar{Q}^A_{\dot\alpha}, S^B_{\alpha} \} }\: ,
\label{cosetdef}
\ee
where the elements of $H$ form the stability group of unbroken
generators.  Although there is a great deal of ambiguity in the choice
of coordinates,  it is convenient to choose the coset representative
\be
V(x^\mu,\eta^A_\alpha, \bar\xi^A_{\dot{\alpha}}, \lambda) =
e^{xP} e^{\eta Q} e^{\bar\xi \bar{S}} e^{\lambda D}\: ,
\label{cosrep}
\ee
with inverse
\be
V^{-1}(x^\mu,\eta^A_\alpha, \bar\xi^A_{\dot{\alpha}}, \lambda)
= e^{-\lambda D} e^{-\bar\xi \bar{S}}e^{-\eta Q}
e^{-xP}\: .
\label{cosinve}
\ee
For simplicity the subscript $0$ has been dropped from the collective
coordinates.  The left invariant 1-form
\be
L = V^{-1} dV = e^{-\lambda} dx P + e^{-\lambda/2}( \bar\xi
dx\cdot\bar\sigma + d\eta ) Q
+ e^{\lambda/2} d\bar\xi
\bar{S} + d\lambda D \: ,
\label{leftdef}
\ee
satisfies the Maurer--Cartan equation
\be
dL - L\wedge L \equiv \left(dL^\Lambda + {1\over 2} L^\Delta \wedge
L^\Sigma \,
f_{\Sigma\Delta}^{\ \  \Lambda}\right)T_\Lambda =0\: ,
\label{maurercart}
\ee
where $f_{\Sigma\Delta}^{\ \  \Lambda}$ are the structure constants
and  $T_\Lambda$ denotes the generators of $SU(2,2|4)$ which divide
into those that are in the coset and those that are in the stability
group,
\be
T_\Lambda = (C_A, \ H_i)\: ,
\label{gendec}
\ee
where $A$ labels the elements of the coset and $i$ the elements of the
stability group.  The one-form may then be decomposed into  the
super-vielbein ($E_M^A$) and $H$-connection ($\omega_M^i$)
\be
L = dZ^M (E_M^A C_A + \omega_M^i H_i) = dZ^M \, L_M^{\
\Lambda}T_\Lambda,
\label{decomps}
\ee
where $M$ is the `coordinate index'.  The components of the
super-vielbein follow from (\ref{leftdef})
\bea
E_\mu^{\hat\mu} &=& e^{-\lambda} \delta_\mu^{\hat\mu}\:, \qquad
E_\mu^{\hat\alpha\hat{A}} = e^{-\lambda/2}( \bar\xi
\bar\sigma_\mu )^{\alpha \hat A}\: ,\qquad
E_{\alpha A}^{\hat\alpha\hat{A}} = e^{-\lambda/2}
\delta_\alpha^{\hat\alpha} \delta^{\hat{A}}_A\: ,\nn \\
E_{\dot\alpha A}^{\hat{\dot{\alpha}}\hat{A}} &=& e^{+\lambda/2}
\delta_{\dot\alpha}^{\hat{\dot{\alpha}}} \delta^{\hat{A}}_A\: ,\qquad
E_d^{\hat{d}} = 1\: ,
\label{supviel}
\eea
with inverse
\bea
E_{\hat\mu}^{\mu} &=& e^{+\lambda} \delta_{\hat\mu}^{\mu}\: , \qquad
E_{\hat\mu}^{\alpha{A}} = - e^{+\lambda}(\bar\xi
\bar\sigma_\mu)^{\alpha A}\: ,
\qquad E_{\hat\alpha\hat{A}}^{\alpha A} = e^{+\lambda/2}
\delta_{\hat\alpha}^{\alpha} \delta^{{A}}_{\hat{A}}\: , \nn\\
E_{\hat{\dot{\alpha}}\hat{A}}^{\dot\alpha A} &=& e^{-\lambda/2}
\delta_{\hat{\dot\alpha}}^{{\dot{\alpha}}} \delta^{{A}}_{\hat{A}}\: ,
\qquad
E_d^{\hat{d}} = 1\: .
\label{invviel}
\eea

The superisometries of the supercoset,
\be
\delta Z^M =  - \Xi^M \: ,
\label{supisom}
\ee
are defined to be those transformations of the super-coordinates that
satisfy
\be
{\cal L}_{\Xi} L + d\Lambda^{(H)}_{\Xi} + [L, \Lambda^{(H)}_{\Xi}] =
\left(\calL_\Xi L^\Lambda + d\Lambda^i\, \delta_i^\Lambda +
\Lambda^i\,
L^\Sigma\, f_{\Sigma i}^{\ \ \Lambda}\right)\, T_\Lambda =0\: ,
\label{supiseq}
\ee
where ${\cal L}_\Xi$ is the Lie super-derivative.  This means that a
coordinate transformation along $\Xi$ can be compensated by a local
$H$-transformation, $\Lambda^{(H)}_{\Xi}$.  This equation is not
$G$-covariant and it is convenient  to rewrite it in terms of a
covariantly constant Killing supervector,
\be
\Sigma \equiv \Sigma^\Lambda G_\Lambda \equiv \Sigma^A C_A + \Sigma^i
H_i\: ,
\label{sigdec}
\ee
which is defined by
\be
\Sigma^{\hat A} = \Xi^A\, E_A^{\ \hat A}\: , \qquad \Sigma^i =
\Lambda^i + \Xi^M\, \omega_M^{\ i}\: .
\label{sigdef}
\ee
By virtue of  (\ref{maurercart}) and (\ref{supiseq}) this satisfies
\be
D\Sigma \equiv d\Sigma + [L, \Sigma] =0 \: .
\label{satsig}
\ee
This equation has the $G$-invariant solution
\be
\Sigma = V^{-1} \Sigma_{_\oplus} V\: ,
\label{killsup}
\ee
where $\Sigma_{_\oplus}$ is any constant element of the Lie algebra
of $G$.
In our case
\be
\Sigma_{_\oplus}^A C_A = x_{_\oplus}^\mu P_\mu +
\eta_{_\oplus}^{\alpha A}
Q_{\alpha A}
+ \bar\xi_{_\oplus}^{\dot{\alpha} A} \bar{S}_{\dot{\alpha} A} +
\lambda_{_\oplus} D\: ,
\label{conssup}
\ee
and
\be
\Sigma_{_\oplus}^i H_i = b_{_\oplus}^\mu K_\mu +
{1\over 2} \omega^{\mu\nu}_{_\oplus} J_{\mu\nu} +
\bar\eta_{_\oplus\,\dot{\alpha} A} \bar Q^{\dot{\alpha} A}
+ \xi_{_\oplus\, \alpha A} S^{\alpha A}\: .
\label{killcomp}
\ee
The explicit expressions for $\Sigma^A$ are
calculated in detail in appendix \ref{supvecs}.
The isometries follow by inverting (\ref{sigdef}),
\be
\Xi^M = \Sigma^A E_A^M = (V^{-1}\Sigma_{_\oplus} V)^A E_A^M\: ,
\label{xifin}
\ee
and these are also given explicitly in appendix  \ref{supvecs}.

\subsection{Grassmann variables for instanton superspace}

The quantities $x^\mu$, $\rho\equiv e^\lambda$, $\eta$ and
$\bar\xi$ are to be identified with the
collective super-coordinates of the instanton. The transformations
 given in appendix B suggest that the fermionic
variables should be packaged together into a sixteen component chiral spinor,
$\Theta^A_a$, where $a=(\alpha,\dot\alpha)$ is a spinor index of
$SO(4,2)$ (or $SO(5,1)$ in euclidean signature). This is achieved by
defining
\be
\Theta_a^A =( \eta^A_\alpha +
x\cdot\sigma_{\alpha\dot\alpha}\bar\xi^{\dot\alpha A},
\bar\xi^{A}_{\dot\alpha})\: .
\label{newthetdef}
\ee
The chirality of this spinor is defined with the chirality of
its four-dimensional spinor components which, in turn, are correlated
with the chirality of the BPST instanton solution.
The 32 supersymmetry parameters are contained in a spinor
$\varepsilon^A_a$ and its  conjugate $\bar\varepsilon^a_A$, defined by
\be
\varepsilon^A_a = (\eta_{_{\oplus}}^A + \sigma\cdot x \,
\bar\xi_{_{\oplus}}^A\, ,\ \  \bar\xi_{_{\oplus}}^A)\: , \qquad
\bar\varepsilon_A^a = (\bar \eta_{_{\oplus}\, A}\, ,\ \   \xi_{_{\oplus}\,
A} +
\sigma\cdot x \, \bar\eta_{_{\oplus}\, A}) \: .
\label{epsdeff}
\ee
The superconformal transformations on the coordinates $x^\mu$, $\rho$,
$\eta$ and $\bar \xi$ can be compactly rewritten as
\bea
\delta \Theta^A_a = \varepsilon^A_a\:
&&\qquad \bar\delta\Theta^A_a = - \Theta^A_b \Theta^B_a
\bar\varepsilon^b_B\: ,
\nn\\
\delta X^M = 0\: , &&\qquad \bar\delta X^M = {1\over 2}
\bar\varepsilon_A
\Gamma^{MN}\Theta^A X_N\: .
\label{superthi}
\eea
The transformations in (\ref{superthi})
are generated by supercharges $\varepsilon^A_a {\cal Q}^a_A$ and $\bar
\varepsilon_A^a
\bar{\cal Q}^A_a$, where
\be
{\cal Q}^a_A = {\partial \over \partial \Theta^A_a},\qquad
\bar{\cal Q}^A_a = \Theta^A_b \Theta^B_a{\partial \over \partial
\Theta^B_b} + {1\over 4}\Gamma^{MN}_a{}^b\Theta^A_b L_{MN}\: ,
\label{supches}
\ee
and satisfy the $SU(2,2|4)$ superalgebra
\be
\{ \calQ^a_A, \bar{\calQ}^B_b\} = {1\over 4}\delta^B_A
\Gamma^{MN}_b{}^a
J_{MN} + {1\over 4}\delta^a_b \widehat{\Gamma}^{ij}_A{}^B T_{ij}\: ,
\label{superss}
\ee
with
$\{\calQ,\calQ\}= \{\bar\calQ, \bar\calQ\}=0$,
where $J_{MN} = L_{MN} + S_{MN}$ are the standard generators of
$SO(4,2)$
and
$T_{ij}$ are the generators of $SO(6)$. More explicitly,
\be
L_{MN} = X_M \partial_N - X_N \partial_M\: , \qquad
S_{MN} ={1\over 2}\,  \Theta^A_a \Gamma_{MN}^a{}_b
{\partial \over \partial \Theta^A_b}\: , \qquad
T_{ij} ={1\over 2}\,  \Theta^A_a \widehat{\Gamma}_{ijA}{}^B
{\partial \over \partial \Theta^B_a}\: .
\label{algrest}
\ee
 In principle, this algebra can be
extended by adding a term $\alpha \Theta$ to $\bar\calQ$.  This
generates an additional term $\alpha\, \delta^a_b \,\delta^B_A$ on the
right-hand side of (\ref{superss}), which is a $U(1)$ central
extension.  The other anticommutators are unaffected.  In what follows
we have only been able to make sense of $\la W\ra$ by choosing $\alpha
=0$ as in (\ref{algrest}).

\subsection{Superinvariants}

The  bosonic invariant of $SO(4,2)$, $R^2 = \eta_{MN} X^M X^N$,
is also invariant
under $SU(2,2|4)$, as is easily seen from the representation defined
by
(\ref{supches}) and (\ref{algrest}).  Given an invariant of the
bosonic subgroup of $OSp(2,2|4)$, such as the quadratic invariant
$U\equiv X_L^2 = (X_4)^2 - (X_1)^2 - (X_2)^2 = R^2 - (X_3)^2 - (X_0)^2
- (X_5)^2\equiv R^2 -V$, it is natural
to ask whether it is possible to find its supersymmetric
extension.

The condition that a quantity $\Psi$ should be invariant under a linear
combination  of the supersymmetries requires
\be
(\varepsilon^A_a \calQ^a_A + \bar\varepsilon^A_a \bar{\calQ}_A^a)
\Psi(X,\Theta) =
0\: ,
\label{inveq}
\ee
with the boundary condition
$\Psi(X,0) = \Psi_B(X)$ for some bosonic
invariant $\Psi_B(X)$. The restriction to $OSp(2,2|4)$ is implemented by
choosing
\be
\bar\varepsilon_A^a = \Omega_{AB} H^{ab} \varepsilon_b^B\: ,
\label{restrinv}
\ee
where $\Omega_{AB} = n_i \widehat{\Gamma}^i_{AB}$ is antisymmetric in
$A$ and $B$ and $H^{ab}=\Gamma_{412}^{ab}$ is symmetric in $a$ and
$b$. The former is a symplectic metric of $Sp(4)\approx SO(5)$ while
the latter is a symmetric metric of $SO(2,2)$. Each  of these can be
used to raise and lower indices of the relevant bosonic subgroup.  The
condition (\ref{restrinv}) eliminates half of the supersymmetry
parameters, so (\ref{inveq}) ensures invariance under a total of
sixteen residual supersymmetries that may be parameterized by
$\varepsilon_a^A$.  In our case the superinvariant, $\Psi(X,\Theta)$, will be
the Wilson loop density in supermoduli space, $W(X,\Theta)$,  while
$\Psi_B(X)$ will be the bosonic density $W_B(X)$ (defined in
(\ref{wbdeff})).  Equation (\ref{inveq}) can be written as
\be
{\partial W(X,\Theta) \over \partial \Theta^A_a} +
\Omega_{AB}H^{ab} \left[
\Theta^B_c \Theta^C_b{\partial W(X,\Theta) \over \partial
\Theta^C_c} +  {1\over 4}\Gamma^{MN}_b{}^c\Theta^B_c L_{MN}\,
W(X,\Theta)
\right] = 0\:
\label{becomw}
\ee with $W(X,0) = W_B(X)$.  It is convenient to rewrite this in the
form
\be
D^{aB}_{Ab}(\Theta) {\partial W(X,\Theta) \over \partial
\Theta^B_b} = - {1\over 4} \Gamma^{MN}_b{}^c\Theta^B_c L_{MN}\,
W(X,\Theta)\: ,
\label{convrew}
\ee
where
\be
D^{aB}_{Ab}(\Theta) = \delta^a_b \delta^B_A +
\Omega_{AC} H^{ac} \Theta^C_b \Theta^B_c\: .
\label{evdef}
\ee
Inverting $D^{aB}_{Ab}(\Theta)$ and setting $B(t) = W(X,t \Theta)$
gives
\bea
{dB(t)\over dt} = {\Theta\over t} {\partial W(X,t\Theta)
\over \partial \Theta} &= &- {t\over 4} \Theta \, D^{-1}(t\Theta) \,
\Omega H  \Gamma^{MN} \Theta \, L_{MN} W(X,t\Theta) \nn\\ &=& -
{1\over 4t} b^{MN}(t\Theta)\, L_{MN}\, W(X, t\Theta)\:,
\label{tineq}
\eea
where
\be
b^{MN}(\Theta) =  \Theta D^{-1}\,(\Theta)\,
\Omega H \Gamma^{MN} \Theta  \: .
\label{ftdef}
\ee
This has a formal solution
\be
B(t) = \calP \exp\left(-\int_0^t
{dt'\over 4t'} \, b^{MN}(t'\Theta)L_{MN} \right) B(0),
\label{solinv}
\ee
so that
\be
W(X,\Theta) = B(1) = \calP\exp\left(-\int_0^1 {dt\over 4t} 
\, b^{MN}(t\Theta)L_{MN} \right) W_B(X)\: .
\label{psisol}
\ee
The symbol $\calP$ in these expressions denotes that the exponentials
are defined by path ordering the operators in the usual manner.  Given
the quantity $W(X,\theta)$, the Wilson loop expectation value is given
by the integral
\be
\la W \ra =  \int d^{16}\Theta
\int {d^6 {X}\over R^4} \delta (X_L^2 - X_T^2 - R^2) W(X,\Theta)\:
\label{wilsfdef}
\ee
(recalling the conventions of (\ref{convens})).

At this point it becomes apparent that fermions enter the Wilson loop
density in a remarkably simple fashion. The density $W(X, \Theta)$ is
simply obtained from the bosonic expression $W_B(X)$ (\ref{wbdeff}) by
the replacement
\be
X^M \rightarrow \tilde{X}^M = R^M{}_N(\Theta) X^N
\label{rotx}
\ee
where the matrix $R(\Theta)$ is given by the six-dimensional
(fundamental) representation of the operator $\calP\exp(-\int dt\, b^{MN}
L_{MN}/4t)$ that acts as a rotation on the $X$ coordinates.  This
observation makes it seem as though the fermionic coordinates can be
eliminated from the integrand simply by changing the integration
variables from $X$ to $\tilde X$.  The jacobian for this change of
variables is unity so that the fermionic variables disappear from the
density and the resulting expression for the Wilson loop expectation
value  can be written as
\be
\la W \ra = \int d^{16}\Theta \int
{d^6\tilde{X}\over R^4} \delta (\tilde{X}_L^2 - \tilde{X}_T^2 - R^2)
W_B(\tilde{X})\: .
\label{genrot}
\ee
The Grassmann integrals apparently vanish.  However, there is an
important subtlety due to the fact that the bosonic integral diverges
and the expression is really of the form $0 \times \infty$.  This
means that it must be regulated by cutting off the region near the
boundary.  We will choose to impose the cutoff $X_4 \le \Lambda$,
where $\Lambda$ is large.  For fixed $|x|$ this translates into a
cutoff $\rho\ge \epsilon$,  where $\epsilon=\ell^2/\Lambda$ is small
(and we have chosen $\ell=R$), so it is cut off at the boundary of
$AdS_5$.  This cutoff is invariant under $SO(5)$ transformations but
it is not Poincar\'e invariant and, as we will discuss in subsection
\ref{altcut},  does not preserve any of the supersymmetries.  In the
presence of this cutoff  the change of variables (\ref{rotx})
introduces a dependence on the fermion coordinates in the
cutoff-dependent endpoint of the bosonic integral.  This suggests a
nonzero result may arise as a boundary term.  Although it seems
probable that the result can be determined by careful analysis
of this boundary term we shall proceed by evaluating the integral
directly in the original coordinates.

In order to evaluate the integral in (\ref{wilsfdef})  we will first
need to expand the exponential in (\ref{psisol}) to select the
sixteenth power of $\Theta$.  This produces a series of powers of $L$
acting on $W_B$. Eventually only the even powers will survive the
bosonic integration  and need to be evaluated.  A great
simplification emerges from observing that
\be
b^{MN}(\Theta) = \Theta D^{-1} \Omega H\Gamma^{MN} \Theta =
\bar\Theta_A
\Gamma^{MN} \Theta^A - \bar\Theta_A \Theta^B \, \bar\Theta_B
\Gamma^{MN} \Theta^A\: ,
\label{simpleth}
\ee
where we are using the short-hand notation
\be
\bar \Theta_A^a \equiv H^{ab}\Omega_{AB}\, \Theta^B_b\: .
\label{barthdef}
\ee
All higher powers of $\Theta$ in
$b^{MN}(\Theta)$  vanish as can be seen by
making use of the antisymmetry of $H^{ab}\, \Theta^A_a \Theta^B_b$
under interchange of the $Sp(4)$ indices $A$ and $B$. An identity
that
needs to be used  here and later on is
\bea
\bar\Theta_A \Theta^B \,
\bar\Theta_B \Gamma^{MN} \Theta^A &=& \Omega_{AD} \Omega_{BC}
\Theta^A
H \Theta^B\,  \Theta^C H \Gamma^{MN} \Theta^D \nn\\ &=& {1\over 3}
\varepsilon_{ABCD} \Theta^A H \Theta^B\,  \Theta^C H \Gamma^{MN}
\Theta^D\: .
\label{idenfour}
\eea
Antisymmetry on the R-symmetry indices also implies that the only
nonzero elements of the terms in (\ref{simpleth}) are those that are
in the coset.  This means that the only nonzero terms arise when $M$
is a `longitudinal' index of $SO(2,1)_L$ and $N$ is a `transverse'
index of $SO(3)_T$.  Therefore only the subset of $L_{MN}$'s that are
in the coset enter into the exponent of (\ref{psisol}).  It is
therefore convenient to decompose the indices under the subgroup $H$
and define
\bea
\Phi &=& \bar \Theta_A\Gamma^{MN}\Theta^A\, L_{MN} =
2 \bar\Theta_A \Gamma^{ir}\Theta^A
L_{ir}\:, \nn\\
{\cal A}& =& \bar\Theta_A \Theta^B\, \bar\Theta_B
\Gamma^{MN}\Theta^A L_{MN} =
2\bar\Theta_A \Theta^B\,  \bar\Theta_B
\Gamma^{ir}\Theta^A L_{ir} \: ,
\label{futdefs}
\eea
where $i=1,2,3$ label $SO(2,1)_L$ and $r=4,5,6$ label $SO(3)_T$.
Since the operators $\Phi$ and  ${\cal A}$ do not commute with each
other it is essential to take care of the path ordering when expanding
the exponential ${\cal P}\exp\left( \int du(\Phi - u\calA)/8 \right)$
(where $u=t^2$).


\section{Integration over the instanton supermoduli}
\label{N4oneinst}

We will now explicitly evaluate the integral in (\ref{wilsfdef}) in
the presence of a cutoff.

\subsection{General properties of the integral}
\label{genprop}

The Grassmann integration selects the terms with sixteen powers of
$\Theta$ that are obtained by expanding (\ref{psisol}), which have
the schematic form
\be
\left({1\over 8!} \Phi^8 - {1\over 6!2} \Phi^6 {\cal A} +
{1\over 4!2!4}\Phi^4 {\cal A}^2 - {1\over 2!3!8} \Phi^2 {\cal A}^3 +
{1\over 4!16} {\cal A}^4\right) \, W_B(X)\: .
\label{expansex}
\ee
This formula suppresses the combinatorics associated with the path
ordering that is nontrivial since the operators $\Phi$ and $\calA$ do
not commute.  Although the structure of the terms in (\ref{expansex}) is
reminiscent of the expansion of the exponential in (\ref{wildef}) in
powers of $\hat \varphi$ and $\hat A$, described in the
introduction, it is significantly simpler.
While (\ref{wildef}) involved path ordering of
matrices in the gauge group that depend on the position around the
loop the expression (\ref{psisol}) does not have this complication.
In other words, the use of superconformal symmetries  has lead to the
abelianization of the bosonic connection.

Using the constraint $X_{L}^{2}-X_{T}^{2}=R^{2}$ (which commutes with
the operator $R^M_{\ \ N}$ defined in (\ref{rotx})) one can think of
$W_B$ as a function of   $|X_{T}|$ only, so that
\be
L_{ir}
W_B(|X_{T}|) = {X_i X_r\over |X_T|} {\partial\over
\partial |X_T|} W_B(|X_T|)\:
\label{lactdef}
\ee
(recalling that the indices $i$ and $r$  are longitudinal and
transverse, respectively).  As we will see below the Grassmann
integration produces a rather complicated tensor that induces all
sorts of contractions of the $L$'s acting on the bosonic invariant
$W_B(X)$. We intend to carry out the Grassmann integration first, so
we will define
\be
F(|X_T|;R) = \int d^{16}\Theta\,
W(X,\Theta)\: .
\label{capfdef}
\ee
The general structure of this function can be expressed as
\bea
F(|X_{T}|;R)&\equiv&  \int d^{16}\Theta \left[{1\over
8!} \Phi^8 + {1\over 4!2!4} \Phi^4 {\cal A}^2 + {1\over 4!16} {\cal
A}^4\right] W_B (|X_T|) \nonumber \\ &=&
\sum_{n=1}^{8}\sum_{k=0}^{\left[{n+2\over 2}\right]} C_{n+2-2k}^{(n)}
|X_T|^{n-2k} R^{2k} {\partial^{(n)} W_B \over \partial |X_T|^{n} }\:
,
\label{aftred}
\eea
where the intermediate equation is a symbolic summary of the expansion
of the exponential in the integrand.  The odd powers of $\calA$ in
(\ref{expansex}) have been dropped since they are odd in $X$ and
do not contribute to the Wilson loop expectation value since they
vanish after integration.  A crucial point is that the
resulting expression $F(|X_{T}|;R)$, being $H$-invariant, can only
depend on the single invariant $X_{T}^{2}$.

The integral (\ref{wilsfdef}) that defines the Wilson loop expectation
value has a divergent contribution from the infinite volume
of the subspace with constant $X_{T}^{2}$, \ie from the infinite volume
of the integral over each $SO(3) \times SO(2,1)$ orbit that comes from
the region close to the $AdS_5$ boundary (near
$\rho=0$ in the original coordinates).  We will regularize such
divergences by introducing a cutoff $X_4 \le \Lambda= R^2/\epsilon$
which breaks the conformal symmetry.
The $X_4$ integral  simply  gives (again ignoring a known overall
coefficient that will be reinstated at the end)
\bea
\la W\ra_\epsilon
&=&  \int^\Lambda {dX_4\over R^4}\int_{X_l^2 + X_T^2 \leq X_4^2 -
R^2}
d^2X_l
d^3X_T \, \delta(X_4^2 - X_l^2 - X_T^2 - R^2)\, F(|X_T|;R) \nn \\
&=& {1\over R^4}\,
\int_{X_l^2 + X_T^2 \leq \Lambda^2 - R^2} {d^2X_l d^3X_T \over
\sqrt{X_l^2 + X_T^2 + R^2}}\, F(|X_T|;R)\: ,
\label{firstint}
\eea
where the subscript $\epsilon$ indicates the presence of the
cutoff $\Lambda= R^2/\epsilon$.
Performing the elementary integrals over  $X_l$ gives
\bea
\la W \ra_\epsilon &=& {2\pi \over R^4}\,
\int_{|X_T|\le \sqrt{\Lambda^2 - R^2}}\, d^3 X_T\, (\Lambda -
\sqrt{X_T^2 + R^2}) \, F(|X_T|;R)\nn\\
&=&
2\pi \int_{|{\bf X}|\le \sqrt{R^2/\epsilon^2 -1}}\, d^3 {\bf X}\,
\left({R\over \epsilon} - \sqrt{{\bf X}^2 +1} \right)\,F(|{\bf
X}|;1)\: ,
\label{secint}
\eea
where the same rescaling has been used as in (\ref{finre}).

The possible divergences of this integral can be analyzed by noting
the following properties of $W_B(X)$ and its derivatives, which arise
in the definition of $F(|X_T|;R)$ in (\ref{aftred}).  Firstly, note
that (\ref{aftred}) is unaltered if $W_B=
 - \cos(\pi |X_T|/ \sqrt{X_T^2 + R^2})$ is replaced by $W_B^{(0)}
\equiv W_B  - 1$ which has asymptotic behaviour for large $|X_T|$
\be
W_B^{(0)} \equiv W_B - 1   \sim - {\pi^2 R^4 \over 8 |X_T|^4}\: .
\label{asymzero}
\ee
Similarly, derivatives of $W_B$ have asymptotic behaviour
\be
W_B^{(n)} = {\partial^{(n)} W_B \over \partial |X_T|^{n} } \sim
(-)^{n+1} {(n+3)! \pi^{2} \over 4! 2} {R^{4} \over |X_T|^{n+4}}\: .
\label{asymn}
\ee
{}From these expressions it follows that $\la W\ra_\epsilon $ is at
most linearly divergent.  The fact that the quartic divergence of the
purely bosonic integral is absent is a consequence of supersymmetry.

Clearly, a linearly divergent term cannot be present in the exact
solution.  Such a term, which has the form $R/\epsilon$ and is
proportional to the circumference of the loop, would
represent a breakdown of conformal invariance.  However, our
calculation  introduced a cutoff in the moduli space integration that
excludes a region close to the loop.  There is therefore a possibility
that we have ignored a singular contribution that arises when the
instanton touches the loop.  Such a term could cancel any apparent
singular behaviour in the integral.  The coefficient of the term
linear in $1/\epsilon$ in (\ref{secint}) is finite and so the linear
divergence arising from the $\Lambda\to \infty$ limit has the form
\be
{2\pi R\over \epsilon}\, \,\int d^3 X_T \, F(|X_T|;R) \equiv   {2\pi
R\over \epsilon}
\,  \calD \:,
\label{lindiv}
\ee
where $\calD$ is a finite coefficient since the integral converges.

The behaviour of the integral can be analyzed in terms of the
coefficients $C_m^{(n)}$ in (\ref{aftred}), noting that
\be
\int_{0}^{\infty} d|X_T|\, |X_T|^{p}\, W_B^{(n)} = 0 \: ,
\label{intlinh}
\ee
for all $p<n$.
After  a change of variables one can express the terms with  $p=n$
and $p=n+2$ in terms of Bessel functions
\bea
\int d|X_T|\, |X_T|^{n+2}\, W_B^{(n)} &=& (-)^{n+1} (n+2)! \int
d|X_T|\,
|X_T|^{2}
W_B^{(0)}\nn\\
& =& (-)^{n+1} n! {\pi^{2}\over 3} [J_{1}(\pi) + \pi J_{0}(\pi)] R,
\label{bessfuno}
\eea
\be
\int d|X_T|\, |X_T|^{n}\, W_B^{(n)}
= (-)^{n+1} n! \int d|X_T|\, W_B^{(0)} = (-)^{n+1} n! \pi^{2} J_{1}(\pi)
R\: .
\label{bessfunt}
\ee
$J_{1}(\pi)$ and  $J_{0}(\pi)$ are incommensurable while the
coefficients $C_{n+2}^{(n)}$ are essentially integers.  This means
that the  coefficient of the linear divergence gets
independent contributions from these two types of terms.  However, the
coefficients turn out to satisfy the identity
\be
\sum_{n=1}^{8}  (-)^{n} n! C_{n}^{(n)} = 0\: ,
\label{twtwocon}
\ee
which eliminates the $p=n$ contribution.  The remaining contribution
is equal to
\be
\calD \equiv \sum_{n=1}^{8}  (-)^{n+1} (n+2)! C_{n+2}^{(n)}\: .
\label{twocon}
\ee
The evaluation of $\calD$
requires extensive computation that will be described in the next
section.

There is also an apparent subleading logarithmic divergence in
(\ref{secint}).  This can be isolated by taking the derivative of
$\la W\ra_\epsilon$ with respect to $\epsilon$.   Since the integrand
vanishes at the upper bound one gets
\be
\epsilon^2 \,{d\la W\ra_\epsilon\over d\epsilon}
\sim \la W\ra_{\rm lin} + \epsilon\, {\la W\ra_{\rm log}} + \ldots\: ,
\label{logdiv}
\ee
where  $\la W\ra_{\rm lin}$ is the coefficient of the linear divergence and
$\la W\ra_{\rm log}$ the coefficient of logarithmic divergence.  These
quantities depend on the polynomials of degree $n+2$ that multiply
$W^{(n)}$ in $F$.  A little algebra shows that the coefficients
satisfy the condition
\be
\la W\ra_{\rm log} = \sum_{n=1}^{8} C_{n+2}^{(n)} (-)^{n} (n+3)! = 0\: ,
\label{wlogs}
\ee
so that the logarithmic divergence vanishes.

To summarize, the integral (\ref{firstint}) gives an expression of
the form
\be
\la W \ra_\epsilon = \calD\, {2\pi R\over \epsilon} + \calF +
O(\epsilon) \: ,
\label{summfir}
\ee
where  $\calF$  is the  finite  integral
\be
\calF \equiv \la W \ra = - {1\over R^4}\,
\int d^3 X_T \,  \sqrt{X_T^2 + R^2} \, F(|X_T|;R)\: .
\label{findefs}
\ee
Although it is not immediately apparent, the integrand is a total
derivative and this integral only gets a contribution from the
boundary at $|X_T| = \infty$.  This is in line with the expectation based on
the original expression for $\langle W \rangle$, which was the
integral of a total divergence.  Performing the integration in detail
and reinstating all the constants that have been dropped up to now
gives the expression
\be
\langle W \rangle  = {g^8_{_{YM}} \over 4! 2^{50} \pi^7} e^{2\pi i
\tau} \sum_{n=2}^{8} \sum_{k=2}^{n} (-)^{n} {(n+3)!\over k+3}
C_{n+2}^{(n)} \: ,
\label{mainres}
\ee
where  $\tau = \vartheta/2\pi + 4\pi i/ g^2_{_{YM}}$
is the complexified Yang--Mills coupling.

In order to convert this  expression into a number we need to
calculate the coefficients $C_n^{(m)}$, which will be the subject of
section \ref{evall}.

\subsection{Evaluation of the integral}
\label{evall}

The sixteen-component Grassmann integrations can be performed by
decomposing the sixteen-component  
$SO(4,2) \times SU(4)$ variable $\Theta_a^A$ 
into two eight-component spinors.   This is 
achieved by choosing a basis in which $\Omega_{12}=-\Omega_{21} =1$
and $\Omega_{34}=-\Omega_{43}=1$ (with all the remaining components
zero).  This will allow us to separate $\Theta_a^A$ 
into two $SO(6,2)$ spinors,
\be
\hat{\theta} = (\Theta^1_a, \Theta^2_a)\:, \qquad
\check{\theta} = (\Theta^3_a, \Theta^4_a)\: .
\label{chatdef}
\ee
The identifications made in these expressions are clarified by
considering  the decomposition $SO(6,2) \to SO(2,2) \times U(1)
\times SU(2)$, which is also a subgroup of $SO(4,2) \times SU(4)$.  
The $SO(6,2)$ spinors  $\hat \theta$ and $\check \theta$ both 
transform as $(2,1,2)_+ \oplus (1,2,2)_-$, where the  notation refers to
the factors in the subgroup (with $\pm $ being the $U(1)$ charge).  
The overall $SO(4,2)\times SU(4)$ chirality determines the chirality
of the $SO(6,2)$ spinors.  The R-symmetry indices $(1,2)$ and $(3,4)$
are doublets of the $SU(2)$ factor and the $SO(4,2)$ chirality is the
same for all components.  Therefore the two $SO(6,2)$ spinors have the
same chirality which is inherited
from the chiralities of $\Theta^A_a$ with respect
to the $SO(6)$ R-symmetry and the
$SO(4,2)$ conformal symmetry\footnote{There is plenty
of scope for a sign error in determining the absolute sign of the
$SO(6,2)$ chirality so we have performed the following calculations
allowing for either sign.}.  
Recall that this chirality originates from the fact
that the BPST instanton is an anti self-dual solution.

From here on we will replace $SO(6,2)$ by $SO(8)$ for notational
convenience\footnote{This makes no difference to the following
discussion and, in any case, we need to make a Wick rotation of one of
the time-like coordinates in order to evaluate the instanton contribution.}.
The spinor bilinears of relevance to
our problem are rewritten in $SO(8)$ notation by using
\be
\Theta^1 H \Gamma^{i r} \Theta^2 = {1\over 2} \hat{\theta}
\gamma^{ir}
\hat{\theta}\, , \qquad \Theta^3 H \Gamma^{i r} \Theta^4 = {1\over 2} \check{\theta}
\gamma^{ir} \check{\theta}   \: ,
\label{compsthh}
\ee
and
\be
\Theta^1 H \Theta^2 = {1\over
2} \hat{\theta} \gamma^{78} \hat{\theta}\,, \qquad \Theta^3 H \Theta^4 =  {1\over
2} \check{\theta} \gamma^{78} \check{\theta} \: ,
\label{compsth}
\ee
where $i=1,2,3$ are the transverse indices, $r=4,5,6$ are the
longitudinal indices and the $SO(8)$  $\gamma$ matrices are
Clebsch--Gordan coefficients that couple the vector ${\bf 8_v}$ to the
two inequivalent spinors, ${\bf 8_c}$ and ${\bf 8_s}$.

With these identifications $\Phi$ can be rewritten in the form
\be
\Phi = \Omega_{AB} \Theta^A
H\,\Gamma^{MN} \Theta^B\, L_{MN}  = 2(\hat\theta \gamma^{ir}
\hat\theta + \check\theta \gamma^{ir} \check\theta) L_{ir}\: .
\label{phieq}
\ee
 After some
manipulations the quantity $\calA$ defined in (\ref{futdefs}) can be rewritten as
\bea
{\cal A} &=&   \frac{8}{3} \left(\Theta^1 H \Theta^2 \, \Theta^3 \Gamma^{MN}
\Theta^4 + \Theta^3 H \Theta^4\,  \Theta^1 \Gamma^{MN} \Theta^2 +
\frac{1}{8} \varepsilon^{PQRSMN}\,  \Theta^1H\Gamma_{PQ}\Theta^2\,
\Theta^3H\Gamma_{RS}\Theta^4 \right) L_{MN}
\nn\\
&=& {4\over 3}\, \left(
\hat\theta\gamma^{ir}\hat\theta\, \check\theta\gamma^{78}\check\theta
+ \hat\theta\gamma^{78}\hat\theta\,
\check\theta\gamma^{ir}\check\theta + {1\over 2}
\varepsilon^{irksmt}\, \hat\theta\gamma_{ks}\hat\theta\, \check\theta
\gamma_{mt}\check\theta \right) L_{ir}\:  .
\label{caladef}
\eea
The signs of the $\varepsilon$ terms in these expressions are 
correlated with the chirality of the instanton solution.  This
sign changes in the case of an anti instanton.  In that case the chirality
of both $SO(8)$ spinors also changes, which changes the signs of
the other terms in (\ref{caladef}).  Therefore, changing from an instanton
to an anti instanton simply reverses the sign of $\calA$, which leaves
the value of $\langle W\rangle$ unaltered since it only receives 
contributions from even powers of $\calA$.

Substituting (\ref{phieq}) and (\ref{caladef}) into (\ref{psisol}) gives
\bea
W(X,\Theta) &\!=\!& {\cal P} \exp \left\{ -\int_0^1 dt^\prime \left(
\frac{t^\prime}{2} \Theta \Omega H \Gamma^{ir} \Theta\,  L_{ir}
+ \frac{{t^\prime}^3}{2} \Theta \Omega H \Theta\,  \Theta \Omega H
\Gamma^{ir} \Theta \, L_{ir} \right) \right\}\,  W_B(X)\nn \\
&\!=\!& {\cal P}  \exp \left\{ -\int_0^1 du \left[ \frac{1}{4} \left(
\hat{\theta} \gamma^{ir} \hat{\theta} + \check{\theta} \gamma^{ir}
\check{\theta} \right)\,  L_{ir}
+  \frac{u}{6} \left( \hat{\theta} \gamma^{78} \hat{\theta}\,
\check{\theta} \gamma^{ir} \check{\theta}  +   \check{\theta}
\gamma^{78} \check{\theta} \, \hat{\theta} \gamma^{ir} \hat{\theta}
\right. \right. \right. \nn\\
&& \left. \left. \left. + {1\over 2}\varepsilon^{jskuir} \hat{\theta}\gamma_{js}
\hat{\theta}\,  \check{\theta} \gamma_{ku} \check{\theta} \right)\,  L_{ir}
\right] \right\}\,  W_B(X) \,
\label{reso8}
\eea
where $u = {t'}^2$.  In order to evaluate the Wilson loop expectation value we need
to extract the $\hat\theta^8 \, \check\theta^8$ term from the expansion of the
exponential, taking account of the path ordering, $\calP$.
The $SO(8)$ Grassmann variables can be integrated out  by using
the standard result,
\be
\int d^8 \hat\theta \hat\theta \gamma^{m_1 n_1} \hat\theta \ldots
\hat\theta \gamma^{m_4 n_4} \hat \theta = \hat t_8^{m_1 n_1 ...
m_4 n_4} \: ,
\label{standint}
\ee
where $m_r,n_r = 1,...8$ and $\hat t_8$ is a standard $SO(8)$-covariant
tensor.   Similarly, the integral over $\check \theta$ can be expressed in
terms of a tensor $\check t_8$. We will be
using the explicit form of $\hat t_8$ and $\check t_8$
given in \cite{gsa} but the range of
the  indices is restricted to the situation in which $m_r = i_r
=1,2,3,7$ and  $n_r = j_r =4,5,6,8$.  Explicitly, the non-vanishing
elements of either of these tensors are \cite{gsa,greengut}
\begin{eqnarray}
t_8^{i_1r_1i_2r_2i_3r_3i_4r_4} &=& {1 \over 2}
\varepsilon^{i_1r_1i_2r_2i_3r_3i_4r_4}\, \label{teightdef} \\
&-& {1\over 2}\left(
\delta^{i_1i_2}\delta^{i_3i_4} \delta^{r_1r_2}\delta^{r_3r_4} +
\delta^{i_1i_3}\delta^{i_2i_4} \delta^{r_1r_3}\delta^{r_2r_4}
+ \delta^{i_1i_4}\delta^{i_2i_3} \delta^{r_1r_4}\delta^{r_2r_3}
\right) \nn \\
&+&{1\over 2} \left(
\delta^{i_1i_2}\delta^{i_3i_4}\delta^{r_2r_3}\delta^{r_1r_4}
+ \delta^{i_1i_2}\delta^{i_3i_4}\delta^{r_2r_4}\delta^{r_1r_3}
+ \delta^{i_1i_4}\delta^{i_2i_3}\delta^{r_2r_4}\delta^{r_1r_3}
\right. \nn \\
&+& \left.
\delta^{i_1i_4}\delta^{i_2i_3}\delta^{r_1r_2}\delta^{r_3r_4}
+ \delta^{i_1i_3}\delta^{i_2i_4}\delta^{r_1r_2}\delta^{r_3r_4}
+ \delta^{i_1i_3}\delta^{i_2i_4}\delta^{r_1r_4}\delta^{r_2r_3}
\right) \: , \nn
\end{eqnarray}
where the sign of the first term is correlated with the $SO(8)$ chirality of
$\hat\theta$ and $\check\theta$.
This $\varepsilon$ tensor arises from terms that include a
$\gamma^{78}$ factor that can arise from (\ref{compsth}) and
(\ref{caladef}).  The three terms in the first parentheses in
(\ref{teightdef}) involve Kronecker deltas that contract the $i$'s in
the same sequence as the $r$'s.  We will refer to these three terms as
`disconnected' contributions. The six terms in the second parentheses
in (\ref{teightdef}) give rise to `connected' contributions.  A useful
shorthand notation is illustrated by considering the contraction
\begin{eqnarray}
2 t_8^{i_1r_1i_2r_2i_3r_3i_4r_4} M_{i_1r_1}\ldots M_{i_4r_4} &=&
-3 {\rm Tr}M^2\, {\rm Tr}M^2 + 6 {\rm Tr}M^4 \nonumber \\
&=& -3 \rule{0pt}{23pt}  \left( \rule{0pt}{14pt} \hspace*{-1pt}
\right)
 \left( \rule{0pt}{14pt} \hspace*{-1pt} \right)
+6 \, \left[ \rule{0pt}{9pt} \hspace*{-0.2pt}
\raisebox{15.3pt}{\rule{18pt}{0.5pt}} \hspace*{-18.1pt}
\raisebox{-10.2pt}{\rule{18pt}{0.5pt}} \right] \: ,
\label{gencon}
\end{eqnarray}
where $M_{ir}$ is an arbitrary matrix with left and right indices in
separate $SO(3)$'s.  The second line of this equation indicates the
contractions diagrammatically.  The vertical and horizontal lines
indicate contractions of the indices.  The coefficients
indicate the number of such terms that occur.  In the case illustrated
above this number is simply an overall multiplicity.  However, in the
application of interest later on the matrix $M_{ir}$ is replaced by
the operator matrix $L_{ir}$, and it will be important to keep track
of the ordering of the indices within each of these different
combinations.  In other words, many different combinations are
subsumed in the notation of (\ref{gencon}).  Furthermore, although the
$\varepsilon$ term in $t_8$ did not contribute to (\ref{gencon}) it 
does contribute to the expressions that enter into the Wilson loop 
calculation.

After the Grassmann integration each of the terms in
(\ref{expansex}) gives rise to the sum of a very large number of
distinct contractions of powers of $L_{ir}$. These are evaluated
by making repeated use of (\ref{lactdef}) which involves a great
deal of computation.  Some of the details are discussed in
appendix \ref{computea}.  In this way the coefficients $C_m^{(n)}$
are calculated and the various contributions to the linear
divergence and the finite parts  determined. The coefficient,
$\calD$,  of the linear divergence in (\ref{summfir}) turns out to
be a nonzero rational number. For the finite part the result 
is
\be
\la W \ra = \mu_{SU(2)} \,{\pi^3 \over 2^{14}\, 3^4 \, 5}\: ,
\label{rescom}
\ee
where  $\mu_{SU(2)}$ is the standard measure for a single
instanton in $\calN =4 $ supersymmetric $SU(2)$ Yang--Mills,
\be
\mu_{SU(2)} = g_{_{YM}}^{8}\,{1\over 2^{34} \pi^{10}}\, e^{2\pi i\tau}  \:  .
\label{su2meas}
\ee
Since the absolute sign of the 
chirality of the $SO(8)$ spinors is difficult to determine, we
note that with the other choice of chirality the result would be
$\mu_{SU(2)} \, \pi^3 2671/(2^{10}\,3^4\cdot 5\cdot 7)$. The relative
simplicity of (\ref{rescom}) suggests that the first choice is the
correct one.

\subsection{Cutoff dependence --- straight line versus circular
loops}
\label{altcut}

We now want to examine whether the result is independent of the cutoff
and consistent with supersymmetry.
Since the $SO(5)$-invariant cutoff already breaks all
the supersymmetries no supersymmetries survive the introduction of the
loop.  However, the presence of the loop in the theory with a
Poincar\'e-invariant cutoff leads to further supersymmetry breaking.
As discussed in section \ref{expect} the condition (\ref{relloop})
(or, equivalently,  (\ref{restrinv})) defines the combinations of
supersymmetries that are preserved in the presence of the loop.  We
argued that the instanton contribution to  a circular loop should give
a finite value that is independent of the cutoff, but that  a straight
line should receive a vanishing contribution.
In this subsection we shall demonstrate that the
expectation value of a circular Wilson loop is independent of the
cutoff procedure  -- more precisely, we will demonstrate that the
result does not depend on whether we use the $SO(5)$-invariant  or
Poincar\'e-invariant cutoff, $\rho\ge \epsilon$.
In each case the result is given by
(\ref{summfir}).  By contrast, we will see that the cutoff $\rho \ge \epsilon$,
that is natural for the straight line in $\bbR^4$, leads to a vanishing
finite part, $\calF_{\rm line} =0$.

It is easy to see that the $SO(5)$-invariant cutoff breaks all the
supersymmetries, even in the absence of the loop\footnote{We are
grateful to Gary Gibbons for the following argument.}.
To see this, recall that a globally defined
Killing vector in $AdS_5$ is timelike, which means that the vector
$\epsilon \gamma^\mu \epsilon'$ (where $\epsilon$ and $\epsilon'$ are
supersymmetry parameters, or global Killing spinors) is timelike.
However, the surface $X_4 = \Lambda$ preserves $SO(5)$, which is
euclidean de Sitter space and has space-like global Killing vectors.
Therefore, none of the $AdS_5$ supersymmetries can be preserved in the
cut-off theory.

The Poincar\'e invariant cutoff $\rho \ge \epsilon$ is adapted to
loops on $\bbR^4$ and is more commonly used in the context of the
AdS/CFT correspondence (see references in \cite{magoo}) and the
specific context of holographic renormalization (\cite{howtogo} and
references therein).  In the absence of the loop this cutoff preserves
the sixteen Poincar\'e supersymmetries, breaking only the sixteen
conformal supersymmetries.  In terms of the constrained
six-dimensional coordinates $\rho = R/(X_{4} - X_{5})$ (recall that we
have set $\ell=R$), so a cutoff at small $\rho$ is equivalent to a cutoff
at large light cone coordinate $X^{-} = X_{4} - X_{5}$.  Whereas the
endpoint $X_4 =  \Lambda = R^2/\epsilon$ intersects the constraint
hyperboloid, $(X_{4})^{2} - (X_{5})^{2} - (X_{0})^{2} - (X_{1})^{2} -
(X_{2})^{2} - (X_{3})^{2} = R^{2}$, on a four-sphere, the endpoint
$X^- = R^2/\epsilon$ intersects the hyperboloid  on a paraboloid.

We will now see that the expressions obtained in earlier sections are
in accord with these symmetry considerations, once a divergent
perimeter term is subtracted.

\subsubsection{Cutoff independence}
\label{subsubone}

Whereas the expression  (\ref{secint}) for a circular loop expectation
value was obtained from (\ref{firstint}) with  $X_4\le R^2/\epsilon$
we here consider the cutoff $X_4 \le  R^2/\epsilon + X_5$.
Integration over the two $X_l$'s followed by the integral over $X_{4}$
gives
\be
\la W \ra'_\epsilon = {1\over R^4}\,
\int d^{2} X_{t}\, dX_5\,  \left({R^2\over \epsilon}+
X_{5} - \sqrt{ X_{T}^{2} +
R^{2}}\right)\,
F(|X_{T}|;R)\: ,
\label{intcut}
\ee
where the ${}'$ indicates the use of the alternative cutoff.  This
expression is similar to the earlier one (\ref{secint}) that used the
$X_4 \le \Lambda$ cutoff, apart from the presence of the term linear
in $X_5$.  The boundary conditions require $R^2/\epsilon + X_{5} \ge
\sqrt{ X_{T}^{2} + R^{2}}$, or
\be
X_5 \ge {\epsilon\over 2R^2}(X_t^2 + R^2) - {R^2\over 2\epsilon}\: .
\label{bounli}
\ee
This means that there is no upper limit to the $X_5$ integral, but
since (\ref{intcut}) converges for large $X_5$ this does not cause a
problem.  The term linear in $X_5$ is antisymmetric under $X_5 \to
-X_5$ so that it is useful to write it as the sum of two
contributions,
\be
\int_{-\tilde\Lambda/2}^\infty dX_5 \, X_5\,
F(|X_{T}|;R)=
\int_{-\tilde\Lambda/2}^{\tilde\Lambda/2} dX_5\, X_5\, F(|X_{T}|;R)
+ \int_{\tilde\Lambda/2}^\infty dX_5\, X_5\, F(|X_{T}|;R)\: ,
\label{xfives}
\ee
where $\tilde \Lambda =  R^2/ 2\epsilon -
\epsilon (X_t^2 + R^2)/2R^2 $.  The first term in (\ref{xfives})
vanishes identically.  Moreover the second term
also vanishes as $1/\tilde\Lambda$ in the $\tilde\Lambda\to \infty$
limit.  But $\tilde \Lambda$ is finite only if $X_t^2 \sim R^4/\epsilon^2$,
in which case the second term vanishes at least as fast as
$|X_t|^{-4}$, so its contribution will vanish after integration over
$X_t$ in (\ref{intcut}). Thus,  the $X_5$  term in (\ref{intcut}) gives a
vanishing contribution.

This means that  the expectation value of the circular Wilson loop,
$\la W \ra'_\epsilon$, has the same value, (\ref{summfir}), that was
found earlier, so the result
is not  sensitive to the cutoff procedure.  We could presumably  have
regularized the calculation by cutting out any small region  of
$AdS_5$ with
topology $B^3 \times  S^1$ around the loop and obtained the same result.

The presence of a linearly divergent term proportional to the
circumference of the loop is obviously inconsistent with conformal
invariance and an artifact of the non supersymmetric
cutoff.  As argued in the
introduction, a more complete treatment would consider the cutoff
theory obtained by the spontaneous symmetry breaking $SU(3) \to SU(2)
\times U(1)$.  In that case there is a dynamical cutoff induced by
fluctuations of the $W$-boson test particle that defines the loop.
This radically changes the behaviour of instantons with scales
smaller than the inverse $W$-boson mass, $M^{-1}$. In the $M\to
\infty$ limit such effects are localized on the loop and should
generate a perimeter effect. Therefore, in the absence of a complete
analysis of such effects, a pragmatic procedure for eliminating the
divergent term is to make a local modification of the Wilson loop by
absorbing the linearly divergent term into a constant
`renormalization' of the test particle mass.

The fact that the terms ignored by our cutoff procedure are localized
on the loop suggests that  the finite result (\ref{rescom}) that
remains after subtraction of the divergent term  is determined
uniquely.  Since it is independent of the radius the calculation  can
be repeated for any value of  $R$,  including the straight
line\footnote{In this argument we are considering general radii with
$R\ne \ell$.}.  In each case the linearly divergent term can be
subtracted by the {\it same} mass `renormalization', with the same
finite result.

\subsubsection{The straight Wilson line}
\label{subsubtwo}

However, as emphasized in the context of the bosonic model of section
\ref{pureYM}, there is a different definition of the straight Wilson
line in $\bbR^4$ which utilizes the Poincar\'e-invariant cutoff.  This is
pathological from the point of view of $S^4$ since both the loop and
the cutoff surface touch the north pole.

 The integrand for the straight line expectation value  can be
obtained from that of the circular loop (\ref{secint}) in the same way
as in the bosonic case where (\ref{singint}) is obtained from
(\ref{finre}).  The result for the straight line per unit length is
simply equal to
\be
\la W({\rm line}) \ra_\epsilon = {1\over \epsilon}\,
\int_{{\bf X} \le \sqrt{R^2/\epsilon^2 -1}}\, d^3{\bf X}\,  F(|{\bf
X}|;1)\: .
\label{secintnew}
\ee
This is a pure linear divergence which vanishes after
subtracting {\it the same} mass term that renders the circular loop
finite. This is in accord with the constraints of supersymmetry discussed in section
\ref{expect}.


\section{Discussion and other issues}
\label{issues}

 In summary, after subtracting a
linearly divergent term, we have found the finite value in (\ref{rescom}) for the
expectation value of a circular
Wilson loop of radius $R$.  The prescription for subtracting
the perimeter
divergence corresponds to the addition of a counterterm for the (infinite) mass
of the test $W$-boson that defines the loop. We believe that the specific finite result
that remains after subtraction of  the divergent term is unambiguous since there is no
 candidate local counterterm that would give a finite contribution.  Importantly, the
 same mass counterterm leads to the vanishing of the instanton contribution
to the straight Wilson line, as required by supersymmetry.
However, understanding the precise origin of this
subtraction is an obvious challenge.
It indicates a contribution that is missing from our cutoff
prescription and arises from small instantons touching the loop.

The problem is that the BPST instanton solution is not an exact
solution of Yang--Mills theory in the presence of the Wilson loop.  We
argued that the deviation from the exact solution should not be
relevant in the semi-classical ($g_{_{YM}} \to 0$) approximation in
most of moduli space.  However, a zero size instanton touching the
loop is a particularly singular configuration that we excluded by our
cutoff.  No matter how small $g_{_{YM}}$ is, in this region of moduli
space the exact solution is very far from the BPST solution.
Therefore, the fact that our cutoff neglects the effect of such
small-scale instantons localized on the loop might be responsible for
the divergent perimeter term.  If, instead, the  theory  were to be
regularized by considering the Wilson loop to be the holonomy of a
$W$-boson of large but finite mass, as described in the introduction,
these singular small-scale instanton contributions should
automatically be incorporated.

The calculations described in this paper raise a number of other
interesting issues that we will now turn to.

\subsection{Generalization to $SU(N)$}

Having evaluated the instanton contribution to the Wilson loop for
gauge group $SU(2)$ we may now consider the extension to $SU(N)$ and
other groups.  The  semi-classical calculation turns out to be very
simple.  The additional bosonic moduli parameterize the coset
$SU(N)/SU(2)\times SU(N-2)\times U(1)$.  Since the Wilson loop is
gauge invariant the extra $4N-8$ bosonic integrations simply give the
volume of the coset.  This amounts to multiplying the $SU(2)$ group
theoretic coefficient with
\be
b_N = {2^{4N-8} \pi^{4N-8} \over (N-1)! (N-2)!}
\left( \rho \over g_{_{YM}} \right)^{4N-8} \; .
\label{bosextra}
\label{groupb}
\ee

Similar considerations apply to other gauge groups.  The extra
fermionic zero modes are a different story. As is well known, only 16
of them -- those associated with the broken supersymmetry and
superconformal transformations -- are exact. The other $8N-16$,
commonly called $\nu^{Af}$ and $\bar\nu^B_f$ ($f=1,...,N-2$)  enter
into the moduli space action as the quartic interaction
\be
S_{4F} = {\pi^2\over 4 g_{_{YM}}^2 \rho^2}\varepsilon_{ABCD}\nu^{Af}
\bar\nu^B_f \nu^{Ch} \bar\nu^D_h\: .
\label{quarticf}
\ee
This means that the Grassmann integration over these variables can be
saturated by bringing down $2N-4$ powers  of $S_{4F}$ from
$e^{-S_{4F}}$.  In fact, this is the leading contribution to the
Wilson loop expectation value for small $g_{_{YM}}$.  Although the
fields $A$ and $\varphi$ in the Wilson loop integrand (\ref{wildef})
can also soak up the extra fermionic variables such contributions are
suppressed by powers of $g_{_{YM}}$.  Consequently, the leading
contribution can be computed by means of a familiar
(Hubbard-Stratonovich) transformation as demonstrated in \cite{dorey},
\be
a_N = \int \prod_{A,f} d\nu^A_f d\bar\nu^B_f\, \left({g_{_{YM}}^2\over
2\pi^2}\right)^{4(N-2)}\,
 e^{-S_{4F}} =
{(2N-2)! \over 2} \left(g_{_{YM}}^2 \over 8\pi^2 \rho^2 \right)^{2N-4}\: .
\label{liftg}
\ee
When combined with the bosonic factor (\ref{bosextra})  this
simply modifies the overall coefficient of the measure but does not
affect its non-trivial dependence on the collective coordinates and on
$g_{_{YM}}$. The result is
\be
\la W \ra_{SU(N)} = {(2N-2)! \over 2^{2N-3} (N-1)! (N-2)!}
\la W \ra_{SU(2)}\: .
\label{resgrp}
\ee

The  generalization to higher instanton numbers $|k|>1$ is more
difficult for general values of $N$.  First of all it is not possible
to use the symmetry arguments that we have used to simplify the
problem.  On top of that we are faced with the problem of integrating
over the unknown multi-instanton moduli space.  As shown in
\cite{dorey} drastic simplifications emerge from a saddle point
evaluation of the large $N$ limit, where duality with multi
D-instanton effects suggests the multi-instanton moduli space
collapses to a copy of $AdS_5$. This means that the dominant region of
moduli space is the one in which all the instantons are at the same
position and have the same scale and lie in commuting $SU(2)$ factors
inside $SU(N)$ \cite{witt,bgkr,dorey}. Once again, in this limit and
in the semi-classical approximation, the non-trivial part of the
computation is already contained in  the $SU(2)$ instanton
calculation.  The overall measure for arbitrary $k$
at leading order in the $1/N$ expansion given in \cite{dorey} replaces
the factor in (\ref{resgrp}).

\subsection{Speculations concerning AdS/CFT}

The really interesting question from the point of view of the AdS/CFT
correspondence is what happens at strong 't Hooft coupling, $\lambda
\equiv g_{_{YM}}^2N\to \infty$?  In this situation it is not possible
to neglect the contributions that arise from the $\nu^{Af}$ and $\bar
\nu^B_f$ modes in the exponent of the one-instanton contribution to the
Wilson loop density.  Roughly
speaking each  fermion bilinear $\bar\nu^{[A}_f\, \nu^{B]f}$ is
replaced by $\sqrt \lambda$ while  $\bar\nu^{\{A}_f\, \nu^{B\}f}$ is
replaced by $g_{_{YM}}$.  This gives plenty of scope for reproducing
the kind of $\exp \sqrt \lambda$ factors that enter into the
perturbative expressions.  However, for large $\lambda$ there are many
other sources of perturbative corrections to the instanton calculation
that also have to be considered, which is a sobering prospect.

Clearly there should be a string theory viewpoint that corresponds to
the Yang--Mills calculations even though we are unable to calculate
the instanton contribution in the limit of large $\lambda$ and make a
direct comparison with type IIB supergravity.  The arguments of
\cite{dg} are in accord with the expectation \cite{mald,rey} that the
Wilson loop in the strongly coupled perturbative sector is  determined
by the functional integral over all world-sheets bounded by the loop
-- the world-sheet of minimal area being the dominant configuration.
The addition of an instanton corresponds to the addition of a
D-instanton in $AdS_5$.  The general supersymmetry considerations
of section \ref{expect} again imply a non-zero contribution to
circular loops.

Making quantitative headway with this description does not appear to
be simple.  However, the strong constraints implied by $SL(2,\bbZ)$
S-duality might help.  This is first seen from the perturbative
formula for the Wilson loop of  \cite{esz} which is in qualitative
accord with the supergravity side of the AdS/CFT correspondence where
the exponential part of the loop expectation value has the form,
$\exp(-A_{\rm min}) = \exp(\ell^2/\alpha') = \exp
{\sqrt{g_{_{YM}}^2N}}$.  This satisfies the simplest of these
conditions
\be
\la W_{1,0}(R;g_s)\ra  \sim \la W_{0,1}(R;1/g_s)\ra \:,
\label{simpcorr}
\ee
where $g_s = g_{_{YM}}^2/4\pi$ and $W_{p,q}$ is the Wilson loop in
which the test charge is a dyon with electric charge $p$ and magnetic
charge $q$.  The symbol $\sim$ indicates that the matching is not
known to be exact when the prefactors multiplying the exponentials are
included.  However, an equality is expected  by Montonen--Olive
duality which equates the expectation value of a Wilson loop for an
electrically charged test particle with coupling constant $g_s$ to
that of a Wilson loop with a magnetically charged test particle (a 't
Hooft loop) with coupling constant $1/g_s$ (when $\vartheta=0$).  This
is just the simplest example of the more general statement of how the
expectation  values should transform under $SL(2,\bbZ)$.  More generally,
there is a separate species of Wilson loop for each of the infinite
number of different possible dyonic test particles.  Under the
$SL(2,\bbZ)$ transformation of the complex coupling,
\be
\tau \to \tau' = {a\tau + b \over c\tau + d}
\label{gentau}
\ee
($ad-bc = 1$ with integer $a,b,c,d$), the expectation value of the
Wilson loop must satisfy
\be
\la W_{p,q} (R; \tau)\ra = \la W_{r,s}(R; \tau')\ra \: ,
\label{dyonloop}
\ee
where the coprime integers $r,s$ are related to  $p,q$ by $SL(2,\bbZ)$ in
the usual manner.  This indicates that the 't Hooft loop and all the
$p,q$ loops with $q\ne 0$ must have nontrivial dependence on
$\tau_1\sim \vartheta/2\pi$.  In particular, under a shift of $\tau_1$
($a=1$, $b=Z$, $c$=0, $d=1$) the loop $W_{p,q}$ transforms into
$W_{p-Z,q}$.  This means that any loop for a test particle carrying a
magnetic charge transforms into a different loop under an integer
shift of $\tau_1$.  The exponential factor $\exp
{\sqrt{g_{_{YM}}^2N}}$ in the expression for the fundamental Wilson
loop has an obvious generalization that has the correct properties,
namely
\be
\la W_{p,q} (R; \tau)\ra \sim P(\tau,\bar\tau)\, \exp\,
(|p+q\tau|\sqrt{g_{_{YM}}^2N}),
\label{gensl}
\ee
where the prefactor $P$ is undetermined (but was proportional to
$\lambda^{-3/4}$ in the limit of large $\lambda$ in the perturbative
sector considered in \cite{esz}). Although in the case of the fundamental
Wilson loop with $p=1$, $q=0$ the exponential factor does not depend
on $\tau_1= \theta/2\pi$, it is difficult to imagine that the
prefactor has no such dependence.  The constraints of $SL(2,\bbZ)$
covariance typically mix perturbative effects with instanton
contributions of the type discussed in this paper.

\subsection{Instanton $n$-point functions in the presence of a Wilson
loop}

In addition to $\la W \ra$ it is of interest to consider 
correlation functions of gauge invariant composite operators in the
background of the Wilson loop.  Such correlation functions encode
detailed information about the expansion of the Wilson loop in terms
of local operators \cite{maldetal,eszope,plefka} and are interesting for a
variety of other reasons.  Among  many possible choices of correlation
functions, let us focus on two that are particularly special.

Firstly, consider the correlation function $\la \Lambda(x_1) \ldots
\Lambda(x_{16}) W\ra$, where $\Lambda\sim \Tr
(F^-_{\mu\nu}\sigma^{\mu\nu}\, \lambda)$ is the composite operator
dual to the supergravity dilatino.  This correlation function is
special because in an instanton background each dilatino contains one
factor of $\zeta$. This means that, as in \cite{bgkr}, all of the
sixteen superconformal collective coordinates have to be absorbed by
the dilatini.  Consequently, to leading order in $g_{_{YM}}$ the
Wilson loop density contains no fermionic coordinates and is given by
$W_B(X_T)$.  Therefore, the correlation is simply an integral over the
bosonic moduli and has the form
\be
\la \Lambda(x_1) \ldots
\Lambda(x_{16}) W\ra = \int {d^4x_0 d\rho\over \rho^5}
\left(\prod_{r=1}^{16}\, K_{7/2}(x_r; x_0, \rho)\right)\,
\cos\left({2\pi R^2 \over
\tilde\rho^2 + R^2} \right) \: ,
\label{corrwil}
\ee
where $\tilde \rho$ is defined in (\ref{invcent}) and
(\ref{solverho}).    Although the integral cannot be performed
explicitly, the result is manifestly finite and almost certainly
nonvanishing for generic positions, $x_r$, of the dilatini (although
singularities arise when these operators touch the loop).

The second correlation function we will discuss is $\la C(x) \, W\ra$,
where the  operator $C(x) = \Tr(F^-)^2$ is dual to the complexified
dilaton $\tau$ and does not absorb any fermionic modes. This means
that the final bosonic integral  has an additional factor of
\be
C(x) = {\rho^4 \over [(x-x_0)^2 + \rho^2]^4}\: ,
\label{ccorr}
\ee
which is simply the classical value of $C(x)$.  The calculation
can be performed in much the same way as in the case of
the pure Wilson loop, leading to
\be
\la C(x) W \ra = \int {d^4x_0 d\rho\over \rho^5} \,
{\rho^4 \over [(x-x_0)^2 + \rho^2]^4} 
\int d^8\eta d^8\bar\xi \, W[x_0, \rho_0; \eta, \bar\xi] \: ,
\label{ccorrres}
\ee
where $W[x_0, \rho_0; \eta, \bar\xi]$ is the Wilson loop density
derived earlier. The integration over the fermionic moduli can be
performed exactly as in the case of the loop with no additional
insertions. Further integration over the insertion point $x$ 
gives the Ward identity 
\be
\int d^4x \la C(x) W \ra = {\partial \over \partial \tau}\la W
\ra\: .
\label{ward}
\ee
Similarly,
\be
\int d^4x \la \bar C(x) W \ra = {\partial \over \partial \bar\tau}\la
W
\ra\: .
\label{wardbar}
\ee
The sum of (\ref{ward}) and (\ref{wardbar}) gives an identity for
$\partial \la W \ra /\partial \vartheta_{YM}$.  In sectors with
nonzero instanton number the left-hand sides of (\ref{ward}) and
(\ref{wardbar}) appear to be very different since $\bar C$ contains
eight fermionic moduli (since $\bar C \sim (\Tr F^{+})^2$) whereas $C$
contains none.  This indicates that $\partial \la W \ra/\partial
\vartheta_{_{YM}}\ne 0$, which would mean that $\la W \ra$ has a
nontrivial dependence on $\vartheta_{_{YM}}$.

Other correlation functions of composite gauge-invariant operators in
a Wilson loop background can be computed in a similar manner, such as
those involving the lowest chiral primary operator $Q^{ij}$.  These
kinds of calculations may reveal interesting information concerning
the operator product expansion of the Wilson loop
\cite{eszope,plefka} and the structure of non-local operators, much as
the operator product expansion of correlation functions
\cite{gb,bkrs,kon,nonwest,afp} has revealed a rich structure of
local scaling operators.


\section*{Acknowledgements}
We are particularly grateful to John Stewart for introducing us to
REDUCE.  We are also grateful to Costas Bachas, Gary Gibbons, Rajesh
Gopakumar, Juan Maldacena, Nick Manton, David Mateos, Hugh Osborn, Eliezer
Rabinovici, Giancarlo Rossi, Augusto Sagnotti, Adam Schwimmer,
Yassen Stanev and Paul Townsend for
useful conversations. MB thanks PPARC for support under its visiting
fellowship scheme and Clare Hall for the pleasant and stimulating
atmosphere during his sabbatical in Cambridge.  The work of SK is
supported by a Marie Curie fellowship.  Partial support of 
EEC contract HPRN-2000-00122 is also
acknowledged.


\appendix


\section{Six-dimensional representation of conformal transformations
of $AdS_5$}
\label{conformal}

In this appendix we will describe the finite $SO(2,2)$ transformations
that map  $(x^\mu,\rho)$ into $(x^{\mu\prime}, \rho')$, where both
points have the same value of the $SO(2,2)$ invariant, $U$.  We will
use a  $6\times 6$ matrix representation of the elements of the group.
In this representation the generators of translations and special
conformal transformations are given by
\be
(P_{\mu})^{M}{}_{N} = \delta_{\mu}^{M} \eta_{N4} + \delta_{\mu}^{M}
\eta_{N5}
+ \eta_{\mu N} \delta^{M}_{4} + \eta_{\mu N} \delta^{M}_{5}
\label{pdirac} \:
\ee
and
\be
(K_{\mu})^{M}{}_{N} = \delta_{\mu}^{M} \eta_{N4} - \delta_{\mu}^{M}
\eta_{N5}
+ \eta_{\mu N} \delta^{M}_{4} - \eta_{\mu N} \delta^{M}_{5} \: .
\label{kdirac}
\ee
The cube of any of these  matrices vanishes,   $P_\mu^3=K_\mu^3=0$,
which makes it very easy to determine the finite transformations.
A finite translation is represented by
\be
(e^{a^{\mu}P_{\mu}})^M{}_N X^N = X^M+ (a^{\mu} P_{\mu})^M{}_N X^N +
{1\over 2} (a^{\mu} a^{\nu} P_{\mu} P_{\nu})^M{}_N X^N\: ,
\label{transop}
\ee
which implies
\be
{x^\mu}^\prime =  x^\mu + a^\mu\:,  \qquad \rho^\prime =\rho\: .
\label{shiftx}
\ee
Similarly, a special conformal transformation is represented by
\be
(e^{b^{\mu}K_{\mu}})^M{}_N X^N = X^M+ (b^{\mu} K_{\mu})^M{}_N X^N +
{1\over 2} (b^{\mu} b^{\nu} K_{\mu} K_{\nu})^M{}_N X^N\: ,
\label{speck}
\ee
which implies
\bea
{x^\mu}^\prime &=& { x^\mu + b^\mu (x^2 + \rho^2)
\over 1 + 2 b\cdot x + b^2 (\rho^2 - x^2) }
\nn\\
\rho^\prime &=& { \rho \over 1 + 2 b\cdot x + b^2 (\rho^2 + x^2) }
\: .
\label{specktr}
\eea
It is important that these transformations preserve the boundary
$\rho=0$ and that they reduce to the usual four-dimensional conformal
transformations on the boundary.

We are interested in the transformations generated by
\be
\Pi^+_l = R\, P_l + {1\over R} K_l\: , \qquad \Pi^-_t = R\, P_t -
{1\over
R} K_t\: .
\label{pistr}
\ee
Using the matrix representation one finds
\be
e^{\alpha\Pi^+} = 1 +
{1\over 2} \Pi^+ \sinh{2\alpha} +
{1\over 4} (\Pi^+)^2 (\cosh{2\alpha} - 1)\: .
\label{pimatr}
\ee
The action on the coordinates is easiest to describe by decomposing
$x^\mu$  into components  parallel $x_\parallel$ to and orthogonal
$x_\perp$ to the vector $\alpha^\mu$,
which gives
\bea
x^{ \prime (+)}_\parallel &=&  x_\parallel +
a \left(1+ {x^2 + \rho^2 \over R^2} \right)
\over
{1 \over 2}\left(1 + \sqrt{1- {4a^2\over R^2}} \right) +
2 {a x_\parallel\over R^2}
+ {1 \over 2R^2}\left(1 - \sqrt{1- {4a^2\over R^2}} \right) (\rho^2 -
x^2)
\nn\\
x^{\prime (+)}_\perp &=& { \sqrt{1- {4a^2\over R^2}} x_\perp \over
{1 \over 2}\left(1 + \sqrt{1- {4a^2\over R^2}} \right) +
2 {a x_\parallel\over R^2}
+ {1 \over 2R^2}\left(1 - \sqrt{1- {4a^2\over R^2}} \right) (\rho^2 -
x^2)
}
\nn\\
\rho^{\prime (+)} &=& {  \sqrt{1- {4a^2\over R^2}} \rho
\over
{1 \over 2}\left(1 + \sqrt{1- {4a^2\over R^2}} \right) +
2 {a x_\parallel\over R^2}
+ {1 \over 2R^2}\left(1 - \sqrt{1- {4a^2\over R^2}} \right) (\rho^2 -
x^2) }\: ,
\label{newsym}
\eea
where
\be
a ={R\over2} \tanh{2\alpha}\: .
\label{adef}
\ee

Similarly using the matrix representation one finds
\be
e^{\beta\Pi^-} = 1 +
{1\over 2} \Pi^- \sin{2\beta}
-{1\over 4} (\Pi^-)^2 (\cos{2\beta} - 1)\: .
\label{pimin}
\ee
The transformation of the coordinates, now decomposed into components
perpendicular and parallel to the vector $\beta^\mu$, is given by
\bea
{x^{ \prime (-)}_\parallel} &=&  {x_\parallel +
b \left(1 -  {x^2 + \rho^2 \over R^2} \right)
\over
{1 \over 2}\left(1 + \sqrt{1 + {4b^2\over R^2}} \right) -
2 {b x_\parallel\over R^2}
+ {1 \over 2R^2}\left(\sqrt{1 + {4b^2\over R^2}} - 1 \right) (\rho^2
- x^2) } \nn\\
x_\perp^{ \prime (-)} &=& { \sqrt{1 + {4b^2\over R^2}} x_\perp \over
{1 \over 2}\left(1 + \sqrt{1 + {4b^2\over R^2}} \right) -
2 {b x_\parallel\over R^2}
+ {1 \over 2R^2}\left(\sqrt{1 +{4b^2\over R^2}} - 1 \right) (\rho^2
-x^2) }
\nn\\
\rho^{\prime (-)} &=& {  \sqrt{1 + {4b^2\over R^2}} \rho
\over
{1 \over 2}\left(1 + \sqrt{1 + {4b^2\over R^2}} \right) -
2 {b x_\parallel\over R^2}
+ {1 \over 2R^2}\left(\sqrt{1+ {4b^2\over R^2}}-1 \right) (\rho^2 -
x^2) } \:,
\label{transpi}
\eea
where
\be
b = {R\over2} \tan{2\beta}\: .
\label{bdefs}
\ee

Since $\Pi_l^{(+)}$ and $\Pi_t^{(-)}$ commute with one another, one
can easily combine the two transformations and  explicitly determine
the parameters  $a^l=a^l_0$ and $b^t=b^t_0$ that give the
transformation taking the instanton to the centre of the loop
$x_0^{\prime} = 0$.  The scale $\tilde \rho$ is thereby determined.


\section{Killing supervectors of the instanton superspace}
\label{supvecs}

In this appendix we shall present the detailed expressions for the
coefficients $\Sigma^A$ that enter into (\ref{sigdec}) and
(\ref{conssup}). We will consider the dependence on the different
constants --- $x_{_\oplus}$, $\lambda_{_\oplus}$, $w_{_\oplus}$,
$b_{_\oplus}$, $\eta_{_\oplus}$, $\bar\eta_{_\oplus}$,
$\bar\xi_{_\oplus}$ and $\xi_{_\oplus}$ --- in turn.

\noindent 1) Translations, $x_{_\oplus}^\mu$:
\be
\Sigma(x_{_\oplus}) =  x_{_\oplus} \left(e^{-\lambda}P +
e^{-\lambda/2}
\bar\xi\sigma Q\right) \: .
\label{transl}
\ee

\noindent 2) Dilations, $\lambda_{_\oplus}$:
\be
\Sigma(\lambda_{_\oplus}) = \lambda_{_\oplus} \left(D - {1\over 2}
e^{+\lambda/2} \bar\xi\bar{S}
+ {1\over 2} e^{-\lambda/2}(\eta +2\bar\xi x\cdot\sigma)Q
+ e^{-\lambda} x\cdot P \right)\: .
\label{dilaf}
\ee

\noindent 3) Rotations, $\omega^{\mu\nu}_{_\oplus}$:
\bea
\Sigma(\omega_{_\oplus}) &=& {1\over 2} \omega^{\mu\nu}_{_\oplus}
\left( J_{\mu\nu} + e^{-\lambda} [x_\mu P_\nu - x_\nu P_\mu]
- {1\over 2} e^{+\lambda/2} \bar\xi \bar\sigma_{\mu\nu}
\bar{S}\right.\nn\\
&&
\left.\qquad\qquad\qquad - {1\over 2} e^{-\lambda/2} \eta
\sigma_{\mu\nu} Q
+ e^{-\lambda/2} \bar\xi \bar\sigma_{\mu}x_{\nu} Q \right)\: .
\label{rotatl}
\eea

\noindent 4) Conformal boots, $b_{_\oplus}^\mu$:
\bea
\Sigma(b_{_\oplus}) &=& b_{_\oplus}^\mu \left(e^{+\lambda} K_\mu -
x_\mu D
-{1\over 2} x^\nu J_{\nu\mu}
e^{-\lambda} [x_\mu x P - {1\over 2} x^2 P_\mu]
+ e^{-\lambda/2} \bar\xi[x_\mu x\sigma - {1\over 2} x^2
\sigma_\mu]Q\right.
\nn \\
&+& \left. e^{+\lambda/2} \left[\eta\sigma_\mu \bar{S}
+ {1\over 2} x_\mu \bar\xi\bar{S}
-{1\over 2} x^\nu \bar\xi \bar\sigma_{\nu\mu} \bar{S}\right]
-{1\over 2} e^{-\lambda/2} \left[x_\mu \eta Q  + x^\nu
\eta\sigma_{\nu\mu}Q\right]\right)\: .
\label{confl}
\eea

\noindent 5) Left supersymmetries, $\eta^{\alpha A}_{_\oplus}$:
\be
\Sigma (\eta_{_\oplus}) =  e^{-\lambda/2} \eta_{_\oplus} Q\: .
\label{leftsusl}
\ee

\noindent 6) Right supersymmetries, $\bar\eta^{\dot\alpha}_{_\oplus
A}$:
\bea
\Sigma (\bar\eta_{_\oplus}) &=&  \bar\eta_{_\oplus}
\left(e^{-\lambda/2}  \bar{Q}
+ e^{-\lambda/2} \bar\sigma\eta \bar\xi\bar\sigma {Q}
+ e^{-\lambda}
\bar\sigma^\mu \eta P
- \bar\xi D - {1\over 4} \bar\sigma^{\mu\nu} \bar\xi J_{\mu\nu}
- {1\over 2} {\hat \Gamma^{ij}}\,  \bar\xi \, T_{ij}\right.
\nn\\
&-& \left. {1\over 4} \bar\xi e^{+\lambda/2} \bar\xi\bar{S}
-{1\over 16} \bar\sigma^{\mu\nu} \bar\xi e^{+\lambda/2}
\bar\xi\bar\sigma_{\mu\nu}\bar{S}+
{1\over 8} {\hat \Gamma}_{ij}\bar\xi e^{+\lambda/2}
\bar\xi {\hat \Gamma}^{ij} \bar{S}\right)\: .
\label{rightsus}
\eea

\noindent 7) Right conformal supersymmetries,
$\bar\xi^{\dot\alpha A}_{_\oplus}$:
\be
\Sigma(\bar\xi_{_\oplus}) = \bar\xi_{_\oplus} (e^{+\lambda/2} \bar{S}
+
e^{-\lambda/2} x\cdot\sigma {Q})\: .
\label{righsussl}
\ee

\noindent 8) Left conformal supersymmetries, $\xi^{\alpha}_{_\oplus
A}$:
\bea
\Sigma (\xi_{_\oplus}) &=& \xi_{_\oplus} \left(e^{+\lambda/2} S +
e^{+\lambda}
\sigma \bar\xi K +  x\cdot \sigma \bar\sigma\eta
\left[ e^{-\lambda} P + e^{-\lambda/2} \bar\xi \bar\sigma
Q\right]\right.
\nn\\
&+&\eta
\left[D+{1\over 4} e^{-\lambda/2} \eta Q- {1\over
4}e^{+\lambda/2}\bar\xi\bar{S}\right]
+{1\over 4} \sigma^{\mu\nu}\eta
\left[J_{\mu\nu}+{1\over 4} e^{-\lambda/2} \eta \sigma_{\mu\nu}
Q\right]
\nn\\
&+&{1\over 2}{\hat \Gamma}^{ij}\eta
\left [T_{ij}+{1\over 4} e^{-\lambda/2} \eta{\hat \Gamma}_{ij} Q
+{1\over 4}e^{+\lambda/2}\bar\xi{\hat \Gamma}_{ij}\bar{S}\right]
+ x\cdot\sigma\bar\xi
\left[D - {1\over 4}e^{+\lambda/2}\bar\xi\bar{S}\right] \nn\\
&+& \left.{1\over 4} x\cdot\sigma\bar\sigma^{\mu\nu}\bar\xi
\left[J_{\mu\nu}+{1\over 4} e^{+\lambda/2} \bar\xi\bar\sigma_{\mu\nu}
\bar{S}\right]
+{1\over 2} x\cdot\sigma \hat\Gamma^{ij}\bar\xi
\left[T_{ij} +{1\over
4}e^{+\lambda/2}\bar\xi{\hat \Gamma}_{ij}\bar{S}\right]\right)\: .
\label{confsups}
\eea

Using these expressions in
(\ref{xifin}) and the expressions for the inverse supervielbeins
(\ref{invviel})  we can deduce the superisometries, as
follows.

\noindent (a) Left supersymmetry:
\be
\delta_{Q} x^\mu = 0 \: ,\qquad
\delta_{Q} \lambda  = 0 \: ,\qquad
\delta_{Q} \eta^{\alpha A} = \eta^{\alpha A}_{_\oplus} \:, \qquad
\delta_{Q} \bar\xi^{\dot\alpha A} = 0\: .
\label{lsusis}
\ee

\noindent (b) Right conformal supersymmetry:
\be
\delta_{\bar{S}} x^\mu = 0\: ,\qquad
\delta_{\bar{S}} \lambda = 0 \: ,\qquad
\delta_{\bar{S}} \eta^{\alpha A} =
\bar\xi^A_{_\oplus\dot\alpha} \bar\sigma^{\dot\alpha \alpha}_\mu
x^\mu\:
,\qquad
\delta_{\bar{S}} \bar\xi^{\dot\alpha A} = \bar\xi^{\dot\alpha
A}_{_\oplus}\: .
\label{rconfis}
\ee

\noindent (c) Right supersymmetry:
\bea
\delta_{\bar{Q}} x^\mu &=& \bar\eta_{_\oplus} \sigma^\mu \eta\:
,\qquad
\delta_{\bar{Q}} \lambda = -\bar\eta_{_\oplus} \bar\xi\: ,\qquad
\delta_{\bar{Q}} \eta^{\alpha A}  = 0\: ,\nn \\
\delta_{\bar{Q}} \bar\xi^{\dot\alpha A} &=&
-{1\over 4} \bar\eta_{_\oplus}\bar\xi\,  \bar\xi^{\dot\alpha A} +
{1\over 16} \bar\eta_{_\oplus} \bar\sigma^{\mu\nu} \bar\xi\,
(\bar\xi\bar\sigma_{\mu\nu})^{\dot\alpha A}
+{1\over 16}  \bar\eta_{_\oplus} {\hat \Gamma}^{ij} \bar\xi \, (\bar\xi
{\hat \Gamma}_{ij})^{\dot\alpha A} \: .
\label{rsusis}
\eea

\noindent (d) Left conformal supersymmetry:
\bea
\delta_{S} x^\mu &=& \xi_{_\oplus} \sigma\cdot{x} \bar\sigma^\mu \eta
+ {1\over 2} e^{2\lambda} \xi_{_\oplus} \sigma^\mu \bar\xi +
\xi_{_\oplus} \sigma_\nu \bar\xi\, (\eta^{\mu\nu} x^2 - 2x^\mu x^\nu)
\: ,\nn  \\
\delta_{S} \lambda &=& -\xi_{_\oplus} (\eta +
\sigma\cdot{x}\bar\xi)\: ,
\nn\\
\delta_{S} \eta^{\alpha A} &=&  -{1\over 4} \xi_{_\oplus} \eta
\eta^{\alpha
A} +
{1\over 16} \xi_{_\oplus}\sigma^{\mu\nu}\eta
\, (\eta\sigma_{\mu\nu})^{\alpha A}
+{1\over 16}  \xi_{_\oplus} {\hat \Gamma}^{ij} \eta \, (\eta {\hat
\Gamma}_{ij})^{\alpha A}\: ,
\nn\\
\delta_{S} \bar\xi^{\dot\alpha A} &=&
-{1\over 4} \xi_{_\oplus} (\eta+\sigma\cdot{x}\bar\xi)
\bar\xi^{\dot\alpha A}
+ {1\over 16} \xi_{_\oplus}{\hat \Gamma}^{ij}(\eta+\sigma\cdot{x}\bar\xi)
\, (\bar\xi{\hat \Gamma}_{ij})^{\dot\alpha A}
\label{lconfis}
\eea

{}From these symmetry transformations  it is simple to deduce the
$OSp(2,2|4)$
transformations generated by (here we are setting $R=1$ for simplicity
of notation) 
\be
G_A = \sigma^{12} Q_A + \Omega_{AB} S^B\: , \qquad
\bar{G}^A = \bar\sigma^{12} \bar{Q}^A + \Omega^{AB} \bar{S}_B \:,
\label{osptwo}
\ee
with constant parameters that we denote by $\epsilon^A_{_\oplus},
\bar\epsilon_{_{\oplus}\, A}$.  These are
\bea
\delta_{G} x^\mu &=&
\Omega_{AB} \epsilon^A_{_\oplus} (\sigma^\mu \bar\xi^B e^{2\lambda} +
\sigma\cdot x \bar\sigma^\mu \eta^B)\: ,
\nn\\
\delta_{G} \lambda &=& \Omega_{AB} \epsilon^A_{_\oplus}
(\eta^B + \sigma\cdot{x}\bar\xi^B)\: ,
\nn\\
\delta_{G} \eta^{A} &=& \sigma^{12} \epsilon^A_{_\oplus}
-{1\over 4} \Omega_{BC} \epsilon^B_{_\oplus} \left(\eta^C  \eta^A +
{1\over 4} \sigma^{\mu\nu}\eta^C (\eta\sigma_{\mu\nu})^A
+  {\hat \Gamma}_{ij\,D}^C \eta^D (\eta {\hat \Gamma}^{ij\, A}\right)\:
,
\nn\\
\delta_{G} \bar\xi^{A} &=&
-{1\over 4}  \Omega_{BC} \epsilon^B_{_\oplus}
\left((\eta^C+\sigma\cdot{x}\bar\xi^C) \bar\xi^{A}
+ 2 {\hat \Gamma}^{ij\, C}_D (\eta^D+\sigma\cdot{x}\bar\xi^D)
(\bar\xi{\hat \Gamma}_{ij})^{A}\right)\: ,
\label{halfind}
\eea
and
\bea
\delta_{\bar{G}} x^\mu &=& \bar\epsilon_{_\oplus A} \bar\sigma^{12}
\sigma^\mu \eta^A \: , \nn\\
\delta_{\bar{G}} \lambda &=& \bar\epsilon_{_\oplus A} \bar\sigma^{12}
\bar\xi^A
\: , \nn \\
\delta_{\bar{G}} \eta^{A} &=&
\Omega^{AB} \bar\epsilon_{_\oplus B} \bar\sigma \cdot{x}  \: ,
\nn \\
\delta_{\bar{G}} \bar\xi^{A} &=&
\Omega^{AB} \bar\epsilon_{_\oplus B}
-{1\over 4} \bar\epsilon_{_\oplus B} \bar\sigma^{12}
\left(\bar\xi^{B} \bar\xi^A
+{1\over 4} \bar\sigma^{\mu\nu}\bar\xi^{B}
(\bar\xi^A\bar\sigma^{\mu\nu}) + {1\over 4} {\hat \Gamma}_{ij\ C}^B
\bar\xi^C \, (\bar\xi^D{\hat \Gamma}^{ij\, A}_D\right) \: .
\label{restl}
\eea

\section{Computation of the integral}
\label{computea}

The various terms that need to be considered in the integrand of the
Wilson loop are those denoted by $\Phi^8$,
$\Phi^4\calA^2$ and $\calA^4$ in (\ref{aftred}).
In this appendix we will sketch
the systematics of the calculation of each of these terms
and the way in which they
contribute to the linearly divergent part of the
integral as well as to the finite part.  The precise
details are too complicated to warrant presentation here (but can be
obtained by direct communication with the authors).

\subsection{The $\Phi^8$ terms}

Since there is no issue of noncommutativity for these terms they are
relatively straightforward to evaluate.  The expansion of eight powers
of $(\hat\theta \gamma^{ir} \hat\theta + \check\theta \gamma^{ir}
\check\theta)$ gives
\be
\Phi^8|_{8\times 8} = \sum_{(70)}
\hat\theta \gamma^{i_1 r_1} \hat\theta ..\hat\theta \gamma^{i_4 r_4}
\hat\theta
\check\theta \gamma^{i_5 r_5} \check\theta .. \check\theta
\gamma^{i_8
r_8} \check\theta L_{i_1r_1}...L_{i_8r_8}\: ,
\label{phieight}
\ee
where the subscript $(70)$ indicates a sum over the $8!/4!4!=70$ terms
that contain four pairs of $\hat \theta$'s and four pairs of
$\check\theta$'s.  Using (\ref{standint}) in order to perform the
integrals over $\hat\theta$ and $\check\theta$ gives a contribution to
the Wilson loop of the form
\bea
\Phi^8 &=& \sum_{(70)}
\hat{t_8}^{i_1r_1..i_4r_4}{t_8}^{i_5r_5..i_8r_8}
L_{i_1r_1}...L_{i_8r_8}\nn\\
&=&
\sum_{(70)} ((6) \delta^{1234} - (3) \delta^{12} \delta^{34})
((6) \delta^{5678} - (3) \delta^{56} \delta^{78}) L_{(1)}...
L_{(8)}\: .
\label{leightcom}
\eea
In this expression the coefficients $(6)$ and $(3)$ indicate the six
connected contributions and the three disconnected contributions that
enter into each of the $t_8$'s.  We have also indicated the index
contractions impressionistically.  Each of the terms  that arises from
(\ref{leightcom}) has the form of a specific  contraction between the
eight powers of $L_{ir}$.  Because each term in the sum over
permutations factorizes into the product of two $t_8$'s the
contractions between the $L$'s also  factorize into two groups --- the
product of contractions on the indices $i_1r_1\ldots i_4r_4$ and the
contractions on $i_5r_5\ldots i_8r_8$.  Performing these contractions
gives a total contribution from the $\Phi^8$ term of the form
\bea
&&{1\over 8!} \Phi^8\, W_B(X) = {1\over 8!} 70\times \left\{ 36 \,
\left[ \rule{0pt}{9pt} \hspace*{-0.2pt}
\raisebox{15.3pt}{\rule{18pt}{0.5pt}} \hspace*{-18.1pt}
\raisebox{-10.2pt}{\rule{18pt}{0.5pt}} \right]
\left[ \rule{0pt}{9pt} \hspace*{-0.2pt}
\raisebox{15.3pt}{\rule{18pt}{0.5pt}} \hspace*{-18.1pt}
\raisebox{-10.2pt}{\rule{18pt}{0.5pt}} \right]
- 36 \,  \left[ \rule{0pt}{9pt} \hspace*{-0.2pt}
\raisebox{15.3pt}{\rule{18pt}{0.5pt}} \hspace*{-18.1pt}
\raisebox{-10.2pt}{\rule{18pt}{0.5pt}} \right]
\left( \rule{0pt}{14pt} \hspace*{-1pt} \right)
\left( \rule{0pt}{14pt} \hspace*{-1pt} \right)
+ 9  \left( \rule{0pt}{14pt} \hspace*{-1pt} \right)
\left( \rule{0pt}{14pt} \hspace*{-1pt} \right)
\left( \rule{0pt}{14pt} \hspace*{-1pt} \right)
\left( \rule{0pt}{14pt} \hspace*{-1pt} \right) \right\} \nn \\
&& = {1\over 8!} \left(L44 - L422 +  L2222\right)
\: .
\label{totleight}
\eea
The notation $L44$ indicates the collection of all terms that are
connected on the  $i_1,r_1\ldots i_4,r_4$ indices as well as on the
$i_5,r_5\ldots i_8,l_8$ indices.  Since the individual $L_{ir}$'s  do
not commute with each other it is important to keep the correct
ordering.  As a result there are $70\times 36 =2520$ distinct terms in
$L44$, corresponding to the different possible contractions.  The
terms in $L422$ are those that are connected on the $i_1,r_1\ldots
i_4,r_4$ indices but disconnected on $i_5,r_5\ldots i_8,r_8$ , as well
as those that are related by interchanging $1,2,3,4$ with
$5,6,7,8$. This also has $2520$ terms.  Likewise $L2222$ indicates
those terms that are disconnected in  both sets of indices.  There are
$70\times 9 = 630$ such terms.

Each of the quantities $L44$, $L422$ and $L2222$ has the form of a
polynomial in $|X_T|$ multiplying a differential of $W_B(X)$, as
indicated in (\ref{aftred}).  Since each of the many thousands of
terms involves the product of eight $L_{ir}$'s there is a substantial
computational problem, for which we have made extensive use of REDUCE
in order to determine the explicit expressions.   Substituting these
in (\ref{secint}) we can extract the $\Phi^8$ contribution to the
linear divergence,
\be
{\calD}1 = {1\over 8!}
\int_0^\infty d|X_T|\, |X_T|^2\, \Phi^8\, W_B(X) =
{1\over 8!} \left(L44 - L422 +  L2222\right) = {35 \over 192}\: .
\label{lindivo}
\ee

Similarly the $\Phi^8$ contribution to the finite term is given by
\be
{\calF}1 = {1\over 8!}
\int_0^\infty d|X_T|\, |X_T|^2\, \sqrt{X_T^2 + R^2}\, \Phi^8\, W_B(X)
= -{1 \over 45} \: .
\label{finphi}
\ee

\subsection{The $\Phi^4 {\cal A}^2$ terms}

The terms $\Phi^4 {\cal A}^2$ need  special attention since $\Phi$ and
${\cal A}$ do not commute. There are 15 distinct orderings of the
$\Phi$'s and ${\cal A}$'s.  In the path ordered expansion of the
exponential in (\ref{psisol}) there is a factor that arises from  the
$u$ integrations that depends on which of these fifteen orderings is
being considered.  Thus, if the two $\calA$ operators are the $p$th
and $q$th positions in the chain of six operators this factor is
proportional to
\be
a_{pq} = \int_0^1 du_1 \int_0^{u_1} du_2 ... \int_0^{u_5} du_6\,
u_p\, u_q\: ,
\label{factorsi}
\ee
where $u_r = t_r^2$.

We will illustrate the procedure for the simple example in which the
two $\calA$'s are the last operators in the chain.  This has the form
\bea
&&\left({4\over 3}\right)^2 
\left((\hat\theta \gamma^{pq} \hat\theta
+ \check\theta \gamma^{pq} \check\theta)L_{pq}\right)^4\,\nn\\
&& \left((2 \hat\theta \gamma^{78} \hat\theta
\check\theta \gamma^{i r}\check\theta + 2 \hat\theta \gamma^{ir}
\hat\theta
\check\theta \gamma^{78}\check\theta +
\varepsilon^{irksmt}
\hat\theta\gamma_{ks}\hat\theta \check\theta \gamma_{mt}\check\theta
)L_{ir}\right)^2\: .
\label{simperm}
\eea
The Grassmann integration selects  the terms with eight $\hat
\theta$'s and eight $\check \theta$'s  in the expansion of this
expression.  It is convenient to group these terms according to the
number of $\varepsilon$ tensors they contain\footnote{We are here
referring to the number of explicit $\varepsilon$'s in the expansion
of (\ref{simperm}). Other factors of $\varepsilon$ arise from the
definition of the $t_8$ tensors.}. The term  with no $\varepsilon$
tensors is
\bea
&& \left({8\over 3}\right)^2
\left\{ \sum_{(6)}
\hat\theta \gamma^{i_1 r_1} \hat\theta \hat\theta \gamma^{i_2
r_2}\hat\theta
(\hat\theta \gamma^{78} \hat\theta)^2
\check\theta \gamma^{i_3 r_3} \check\theta \ldots \check\theta
\gamma^{i_6
r_6} \check\theta + (\hat{\theta} \leftrightarrow \check{\theta})
\right.
\nn\\
&+&\left. \sum_{(6)}
\hat\theta \gamma^{i_1 r_1} \hat\theta 
\hat\theta \gamma^{i_2 r_2} \hat\theta
\hat\theta \gamma^{i_3 r_3}\hat\theta
\hat\theta \gamma^{78} \hat\theta
\check\theta \gamma^{i_4 r_4} \check\theta 
\check\theta \gamma^{i_5 r_5} \check\theta
\check\theta \gamma^{i_6r_6} \check\theta 
\check\theta \gamma^{78} \check\theta \right\}
L_{i_1r_1} \ldots L_{i_6r_6}\: ,
\label{noneps}
\eea
where the sums are over the $6=4!/2!2!$ distributions of 4 elements
into two groups of 2.  Integrating over $\hat\theta$ and
$\check\theta$ gives
\bea
&&\left({8\over 3}\right)^2
\sum_{(6)} ({t}^{i_1r_1i_2r_27878}{t}^{i_3r_3..i_6r_6}
+ {t}^{i_1r_1..i_4r_4}{t}^{i_5r_5i_6r_6 7878}
+ 2 {t}^{i_1r_1.i_3r_378}{t}^{i_4r_4.i_6r_678})
L_{i_1r_1} \ldots L_{i_6r_6}
\nn\\
&& = \left({8\over 3}\right)^2  
\left[6 (- \delta^{12}) ((6) \delta^{3456} - (3)
\delta^{34}\delta^{56})
+ 6 ((6) \delta^{1234} - (3) \delta^{12}\delta^{34}) (-
\delta^{56})\right.
\nn\\
&&+
\left.  12( (6) \delta^{12}\delta^{34}\delta^{56} + (12)
\delta^{123456}
- (18)  \delta^{1234}\delta^{56})\right]L_{(1)}\ldots L_{(6)}\: .
\label{morsix}
\eea
As before, the notation is symbolic, the numbers in parentheses
indicating the number of distinct permutations involved.  The
expression (\ref{morsix}) is again evaluated by a REDUCE programme.  The
terms with one or two $\varepsilon$'s in the expansion of
(\ref{simperm}) must also be included by a similar analysis in order
to complete the first of the fifteen permutations. The result is a
contribution to the $\Phi^4\calA^2$ term that is a sum of terms with
the structures $L6$, $L42$ and $L222$ --- in the earlier terminology
these are connected, partially connected and disconnected,
respectively.

A similar procedure is carried out for each of the other fourteen
distinct orderings of the $\Phi$ and $\calA$ operators in the chain.
They each give rise to contributions to the $\Phi^4\calA^2$ term that
have the structures $L6$, $L42$ and $L222$.  The result of the REDUCE
computation of the sum of these terms gives the contribution to the
linearly divergent part, (\ref{secint}),
\be
{\calD}2 = {1\over 4!2!}
\int_0^\infty d|X_T|\, |X_T|^2\, \Phi^4\calA^2\, W_B(X)
= - \frac{7}{10} \: .
\label{lsixt}
\ee
The contribution to the finite part is
\be
{\calF}2 = {1\over 4!2!}
\int_0^\infty d|X_T|\, |X_T|^2\, \sqrt{X_T^2 + R^2}\, \Phi^4\calA^2\,
W_B(X)
= {11\over 5} \: .
\label{finphia}
\ee

\subsection{The ${\cal A}^4$ terms}

The last set of terms that arises in the expansion of the integrand
are those of the form  ${\cal A}^4$ terms for which there is again no
problem with noncommutativity.

It is again convenient to group terms according to the number of
$\varepsilon$ tensors.  The terms with no $\varepsilon$ come from the
expansion of
\be
\left((\hat\theta\gamma^{ir}\hat\theta
\check\theta\gamma^{78}\check\theta
+\hat\theta\gamma^{ir}\hat\theta \check\theta\gamma^{ir}\check\theta)
L_{ir}\right)^4
= \sum_{(70)}
\hat\theta \gamma^{i_1 r_1} \hat\theta ..\hat\theta \gamma^{i_4 r_4}
\hat\theta
\check\theta \gamma^{i_5 r_5} \check\theta .. \check\theta
\gamma^{i_8
r_8} \check\theta L_{i_1r_1}\ldots L_{i_8r_8}\:,
\label{noepsf}
\ee
where the sum is over the $70=8!/4!4!$ distributions of $8$ elements
into two groups of $4$.  Integrating over $\hat\theta$ and
$\check\theta$ gives
\bea
&&\sum_{(70)} {t}^{i_1r_1..i_4r_4} {t}^{i_5r_5..i_8r_8}
L_{i_1r_1}\ldots L_{i_8r_8}\nn\\
&&=
\sum_{(70)} ((6) \delta^{1234} - (3) \delta^{12} \delta^{34})
((6) \delta^{5678} - (3) \delta^{56} \delta^{78})L_{(1)}\ldots
L_{(4)}\:.
\label{intgf}
\eea
Performing the contractions gives the expression
\be
\calA^4\, W_B(X) = [6L4 - 4 L22]
\: .
\label{totlfour}
\ee

Similar manipulations are needed to determine the contributions from
terms with one, two, three or four factors of $\varepsilon$ in the
expansion of the $\calA^4$ term.  These are to be added together to
get the total contribution.  The result is that these terms give a
contribution to the  linear divergence of the form
\be
{\calD}3 = {1\over 4!}
\int_0^\infty d|X_T|\, |X_T|^2\, \calA^4\, W_B(X) = \frac{80}{27} \:,
\label{finas}
\ee
while the finite contribution from these terms is
\be
{\calF}3 = {1\over 4!}
\int_0^\infty d|X_T|\, |X_T|^2\, \sqrt{X_T^2 + R^2}\,\calA^4 \, W_B(X)
= -{176\over 81} \: .
\label{fina}
\ee

The total coefficient of the linear divergence arising from the sum of
all terms is
\be
\calD = {\calD}1+{\calD}2+{\calD}3  = \frac{21127}{8640} \:,
\label{lintot}
\ee
and the total finite result is
\be
\calF = {\calF}1 + {\calF}2 + {\calF}3 =  \frac{2}{405}\: .
\label{fintoty}
\ee



\begin{thebibliography}{11}


\bibitem{magoo}
O.~Aharony, S.~S.~Gubser, J.~Maldacena, H.~Ooguri and Y.~Oz,
``Large N field theories, string theory and gravity'',
Phys.\ Rept.\  {\bf 323}, 183 (2000)
[hep-th/9905111].

\bibitem{dan}
D.~Z.~Freedman and P.~Henry-Labordere,
``Field theory insight from the AdS/CFT correspondence'',
hep-th/0011086.

\bibitem{mb}
M.~Bianchi,
``(Non-)perturbative tests of the AdS/CFT correspondence'',
Nucl.\ Phys.\ Proc.\ Suppl.\  {\bf 102}, 56 (2001)
[hep-th/0103112].

\bibitem{dfedh}
{E.~D'Hoker and D.~Z.~Freedman,
``Supersymmetric Gauge Theories and the AdS/CFT Correspondence'',
hep-th/0201253.}

\bibitem{dgo}{N.~Drukker, D.~J.~Gross and H.~Ooguri,
``Wilson loops and minimal surfaces'',
Phys.\ Rev.\ D {\bf 60} (1999) 125006
[hep-th/9904191];
H.~Ooguri, ``Wilson loops in large N theories'',
Class.\ Quant.\ Grav.\  {\bf 17} (2000) 1225
[hep-th/9909040].}

\bibitem{mald}{J.~Maldacena,
``Wilson loops in large N field theories'',
Phys.\ Rev.\ Lett.\  {\bf 80} (1998) 4859
[hep-th/9803002].}

\bibitem{rey}
{S.~Rey and J.~Yee,
``Macroscopic strings as heavy quarks in large N gauge theory and 
anti-de Sitter supergravity'', hep-th/9803001.}

\bibitem{graham}
C.~R.~Graham,
``Volume and area renormalizations for conformally compact Einstein  
metrics'', [math.dg/9909042].

\bibitem{orrt}
H.~Ooguri, J.~Rahmfeld, H.~Robins and J.~Tannenhauser,
``Holography in superspace'',
JHEP {\bf 0007} (2000) 045
[hep-th/0007104].

\bibitem{esz}{J.~K.~Erickson, G.~W.~Semenoff and K.~Zarembo,
``Wilson loops in N = 4 supersymmetric Yang-Mills theory'',
Nucl.  Phys.\ B {\bf 582} (2000) 155
[hep-th/0003055].}

\bibitem{dg}{N. Drukker and D.J. Gross, ``An Exact Prediction of N=4
SUSYM Theory for String Theory'', hep-th/0010274.}

\bibitem{ad}G.~Akemann and P.~H.~Damgaard,
``Wilson loops in N = 4 supersymmetric Yang-Mills theory from random
matrix theory'',
hep-th/0101225.

\bibitem{plefka}
{J.~Plefka and M.~Staudacher,
``Two loops to two loops in N = 4 supersymmetric Yang-Mills theory'',
JHEP {\bf 0109}, 031 (2001) [hep-th/0108182]. 
G.~Arutyunov, J.~Plefka and M.~Staudacher,
``Limiting Geometries of Two Circular Maldacena-Wilson Loop
Operators'', hep-th/0111290.}

\bibitem{bgk}
M.~Bianchi, M.~B.~Green and  S.~Kovacs,
``Instantons and  BPS Wilson loops''
DAMTP-2001-57, hep-th/0107028.

\bibitem{dirac} {P.A.M. Dirac, ``Wave equations in conformal space'',
Ann. of Maths. {\bf 37} 429 (1935) 429.}

\bibitem{bg}{T.~Banks and M.~B.~Green,
``Non-perturbative effects in AdS(5) x S**5 string theory and
d = 4 SUSY  Yang-Mills'',
JHEP {\bf 9805} (1998) 002
[hep-th/9804170].}

\bibitem{bgkr}
M.~Bianchi, M.~B.~Green, S.~Kovacs and G.~Rossi,
``Instantons in supersymmetric Yang-Mills and D-instantons in IIB
superstring theory'',
JHEP {\bf 9808} (1998) 013
[hep-th/9807033].

\bibitem{bkrs}
M.~Bianchi, S.~Kovacs, G.~Rossi and Y.~S.~Stanev,
``On the logarithmic behavior in N = 4 SYM theory'',
JHEP {\bf 9908}, 020 (1999)
[hep-th/9906188].

\bibitem{ht}{A.~Hart and M.~Teper,
``Instantons and Monopoles in the Maximally Abelian Gauge'',
Phys.\ Lett.\ B {\bf 371}, 261 (1996)
[hep-lat/9511016].}

\bibitem{Browera}
R.~C.~Brower, K.~N.~Orginos and C.~Tan,
``Magnetic monopole loop for the Yang-Mills instanton'',
Phys.\ Rev.\ D {\bf 55} (1997) 6313
[hep-th/9610101].

\bibitem{Browerb}
R.~C.~Brower, K.~N.~Orginos and C.~I.~Tan,
``Instantons in the maximally Abelian gauge'',
Nucl.\ Phys.\ Proc.\ Suppl.\  {\bf 53} (1997) 488
[hep-lat/9608012].

\bibitem{tHooft}
G.~'t Hooft,
``Computation Of The Quantum Effects Due To A Four-Dimensional
Pseudoparticle'', Phys.\ Rev.\ D {\bf 14}, 3432 (1976)
[Erratum-ibid.\ D {\bf 18}, 2199 (1976)].

\bibitem{supcos}
P.~Claus and R.~Kallosh,
``Superisometries of the AdS x S superspace'',
JHEP {\bf 9903} (1999) 014 [hep-th/9812087];
P.~Claus, J.~Rahmfeld, H.~Robins, J.~Tannenhauser and Y.~Zunger,
``Isometries in anti-de Sitter and conformal superspaces'',
JHEP {\bf 0007} (2000) 047 [hep-th/0007099].

\bibitem{gsa}{M.~B.~Green and J.~H.~Schwarz,
``Supersymmetrical Dual String Theory. 2. Vertices And Trees'',
Nucl.\ Phys.\ B {\bf 198} (1982) 252.}

\bibitem{greengut}
M.~B.~Green and M.~Gutperle, ``Effects of D-instantons'',
Nucl.\ Phys.\ B {\bf 498}, 195 (1997) [hep-th/9701093].

\bibitem{howtogo}
{M.~Bianchi, D.~Z.~Freedman and K.~Skenderis,
``How to go with an RG flow'',
JHEP {\bf 0108}, 041 (2001)
[hep-th/0105276]; ``Holographic renormalization'', hep-th/0112119.}

\bibitem{dorey}{N.~Dorey, T.~J.~Hollowood, V.~V.~Khoze, M.~P.~Mattis
and S.~Vandoren,
``Multi-instanton calculus and the AdS/CFT correspondence in N = 4
superconformal field theory'',
Nucl.\ Phys.\ B {\bf 552} (1999) 88
[hep-th/9901128].}

\bibitem{witt}
E.~Witten,
``Baryons and branes in anti de Sitter space'',
JHEP {\bf 9807}, 006 (1998)
[hep-th/9805112].

\bibitem{maldetal}
D.~Berenstein, R.~Corrado, W.~Fischler and J.~Maldacena,
``The operator product expansion for Wilson loops and surfaces in the
large N limit'',
Phys.\ Rev.\ D {\bf 59}, 105023 (1999)
[hep-th/9809188].

\bibitem{eszope}
G.~W.~Semenoff and K.~Zarembo,
``More exact predictions of SUSYM for string theory'',
Nucl.\ Phys.\ B {\bf 616}, 34 (2001)
[hep-th/0106015].

\bibitem{gb}
J.~H.~Brodie and M.~Gutperle,
``String corrections to four point functions in the AdS/CFT  correspondence'',
Phys.\ Lett.\ B {\bf 445}, 296 (1999)
[hep-th/9809067].

\bibitem{kon}
M.~Bianchi, S.~Kovacs, G.~Rossi and Y.~S.~Stanev,
``Properties of the Konishi multiplet in N = 4 SYM theory'',
JHEP {\bf 0105}, 042 (2001) [hep-th/0104016].

\bibitem{nonwest}
G.~Arutyunov, B.~Eden, A.~C.~Petkou and E.~Sokatchev,
``Exceptional non-renormalization properties and OPE analysis of chiral
four-point functions in N = 4 SYM(4)'', hep-th/0103230. 
G.~Arutyunov, B.~Eden and E.~Sokatchev,
``On non-renormalization and OPE in superconformal field theories'',
Nucl.\ Phys.\ B {\bf 619}, 359 (2001) [hep-th/0105254].
B.~Eden and E.~Sokatchev,
``On the OPE of 1/2 BPS short operators in N = 4 SCFT(4)'',
Nucl.\ Phys.\ B {\bf 618}, 259 (2001)
[hep-th/0106249].

\bibitem{afp}
G.~Arutyunov, S.~Frolov and A.~Petkou,
``Operator product expansion of the lowest weight CPOs in N = 4  SYM(4) at
strong coupling'', Nucl.\ Phys.\ B {\bf 586}, 547 (2000)
[Erratum-ibid.\ B {\bf 609}, 539 (2000)] [hep-th/0005182];
``Perturbative and instanton corrections to the OPE of CPOs in N = 4  
SYM(4)'', Nucl.\ Phys.\ B {\bf 602}, 238 (2001)
[Erratum-ibid.\ B {\bf 609}, 540 (2001)] [hep-th/0010137].


\end{thebibliography}
\end{document}